\documentclass[twocolumn]{aastex63}

\usepackage{graphicx}
\usepackage{ulem}
\usepackage{xcolor}
\usepackage{natbib}
\usepackage{amsmath}
\usepackage{float}
\usepackage{hyperref}
\usepackage{url}
\usepackage{soul}
\usepackage{enumitem}
\usepackage{lineno}
\usepackage{longtable}

\newcounter{rtask}
\newcommand{\rtask}[1]{\refstepcounter{rtask}\label{#1}}

\begin{document}

\title{\large Substructures in Compact Disks of the Taurus Star-forming Region}

\author[0000-0002-8537-9114]{Shangjia Zhang}
\affiliation{Department of Physics and Astronomy, University of Nevada, Las Vegas, Las Vegas, NV 89154, USA}
\affiliation{Nevada Center for Astrophysics, University of Nevada, Las Vegas, Las Vegas, NV 89154, USA}
\correspondingauthor{Shangjia Zhang}
\email{shangjia.zhang@unlv.edu}

\author[0000-0003-2690-6241]{Matt Kalscheur}
\affiliation{Department of Astronomy, University of Wisconsin-Madison, Madison, WI 53706, USA}

\author[0000-0002-7607-719X]{Feng Long}
\altaffiliation{NASA Hubble Fellowship Program Sagan Fellow}
\affiliation{Harvard-Smithsonian Center for Astrophysics, Cambridge, MA 02138, USA}
\affiliation{Lunar and Planetary Laboratory, University of Arizona, Tucson, AZ 85721, USA}

\author[0000-0002-0661-7517]{Ke Zhang}
\affiliation{Department of Astronomy, University of Wisconsin-Madison, Madison, WI 53706, USA}

\author[0000-0003-3840-7490]{Deryl E. Long}
\affiliation{Department of Astronomy, University of Virginia, Charlottesville, VA 22904, USA}

\author[0000-0003-4179-6394]{Edwin A. Bergin}
\affiliation{Department of Astronomy, University of Michigan, Ann Arbor, MI 48109, USA}

\author[0000-0003-3616-6822]{Zhaohuan Zhu}
\affiliation{Department of Physics and Astronomy, University of Nevada, Las Vegas, Las Vegas, NV 89154, USA}
\affiliation{Nevada Center for Astrophysics, University of Nevada, Las Vegas, Las Vegas, NV 89154, USA}

\author[0000-0002-8623-9703]{Leon Trapman}
\affiliation{Department of Astronomy, University of Wisconsin-Madison, Madison, WI 53706, USA}

\begin{abstract}

Observations of substructure in protoplanetary disks have largely been limited to the brightest and largest disks, excluding the abundant population of compact disks which are likely sites of planet formation.  Here, we reanalyze $\thicksim$0.1$\arcsec$, 1.33 mm ALMA continuum observations of 12 compact protoplanetary disks in the Taurus star-forming region.  By fitting visibilities directly, we identify substructures in 6 of the 12 compact disks. We then compare the substructures identified in the full Taurus sample of 24 disks in single star systems and the ALMA DSHARP survey, differentiating between compact ($R\textsubscript{eff,90\%}$ $<$ 50 au) and extended ($R\textsubscript{eff,90\%}$ $\ge$ 50 au) disk sources.  We find that substructures are detected at nearly all radii in both small and large disks.
%, with no obvious correlation between gap width and effective radius of the disk after correcting for gap location. 
 Tentatively, we find fewer wide gaps in intermediate-sized disks with $R\textsubscript{eff,90\%}$ between 30 and 90 au. We perform a series of planet-disk interaction simulations to constrain the sensitivity of our visibility-fitting approach. Under an assumption of planet-disk interaction, we use the gap widths and common disk parameters to calculate potential planet masses within the Taurus sample.  We find that the young planet occurrence rate peaks near Neptune masses, similar to the DSHARP sample.  For 0.01 $M_J/M_\odot$ $\lesssim$ $M_p/M_*$ $\lesssim$ 0.1 $M_J/M_\odot$, the rate is 17.4$\pm$8.3\%; for 0.1 $M_J/M_\odot$ $\lesssim$ $M_p/M_*$ $\lesssim$ 1 $M_J/M_\odot$, it is 27.8$\pm$8.3\%. Both of them are consistent with microlensing surveys.  For gas giants more massive than 5 $M_J$, the occurrence rate is 4.2$\pm$4.2\%, consistent with direct imaging surveys.

\keywords{planets and satellites: detection, protoplanetary disks, planet–disk interactions, circumstellar matter}

\end{abstract}

% SZ: Comments on the abstract. (When the gap width is normalized by the gap location, the inner disk and outer disk have similar gap widths.) With the new substructures found in compact disks, we can do more complete statistics on the Taurus sample. We calculate the potential planet masses using the gap width. We study the young planet occurrence rate for both Taurus and DSHARP samples and find great similarities. The occurrence rate peaks around Neptune masses. For the Taurus sample it is 17.4$_{-8.3}^{+8.3}$\%, for 0.01 $M_J/M_\odot$ $\lesssim$ $M_p/M_*$ $\lesssim$ 0.1 $M_J/M_\odot$, consistent with microlensing surveys.
% The occurrence is 20.8$_{-8.3}^{+8.3}$\%, for 0.1 $M_J/M_\odot$ $\lesssim$ $M_p/M_*$ $\lesssim$ 1 $M_J/M_\odot$, higher than the microlensing surveys.
% The occurrence rate for gas giants $>$ 5$M_J$ is 4.2$^{+4.2}_{-4.2}$\%,, consistent with direct imaging surveys.

\section{Introduction}

One of the most exciting discoveries in recent years is the prevalence of small-scale substructure (e.g., gaps, rings and spirals) in submillimeter/millimeter continuum emission (e.g., \citealt{ALMA_2015}; \citealt{Zhang_2016}; \citealt{Andrews_2016}; \citealt{Long_2018}; \citealt{Huang_2018}; \citealt{Cieza_2021}) and scattered light observations (e.g., \citealt{van_Boekel_2017}; \citealt{Avenhaus_2018}; \citealt{Garufi_2018}) of protoplanetary disks.  Substructures in disks represent a likely solution to the long-standing problems of rapid radial drift (see \citealt{Whipple_1972}; \citealt{Weidenschilling_1997}; \citealt{Takeuchi_2002}) and planetesimal formation (see \citealt{Takeuchi_2005}) inherent to the standard assumption of a smooth gas disk.

A number of physical mechanisms have been proposed to explain such substructures: MHD/photoevaporative winds (e.g., \citealt{Alexander_2014}; \citealt{Suzuki_2016}), zonal flows tied to concentrations of magnetic flux (e.g., \citealt{Johansen_2009}; \citealt{Bai_2014}; \citealt{Suriano_2017}), a global gravitational instability driven by envelope infall that could produce large-amplitude spiral density waves (\citealt{Lesur_2015}) or gaps and rings (e.g., \citealt{Kuznetsova_2022}), changes in dust properties at condensation fronts (\citealt{Zhang_2015}; \citealt{Okuzumi_2016}; \citealt{Pinilla_2017}), and embedded planets (e.g., \citealt{Dong_2015}; \citealt{Zhang_2018}).

However, despite the ubiquity of substructures seen in recent observations and the number of proprosed mechanisms for their generation, studies have been mostly limited to large and bright disks most readily observable at current spatial resolutions (e.g., \citealt{Huang_2018}; \citealt{Cieza_2021}).  The majority of nearby protoplanetary disks are both fainter and more compact (\citealt{Ansdell_2016}; \citealt{Pascucci_2016}; \citealt{Cieza_2019}; \citealt{Williams_2019}).  If substructure is essential to the planetesimal formation problem, and compact disks are likely sites of planet formation, we expect substructures to be a common component of compact disks as well.  If they are due to planets, we can infer planet masses and make a more direct comparison to the planet population within the Solar System than is possible when planet masses are inferred from the inner region of extended disks.

% SZ: If substructures are common in compact disks and due to the planets. We can infer the planet masses in the inner disk ($\sim$ 30 au) and compare the planet population more similar to our Solar System than what inferred from extended disks $\sim$ 100 au. The substructures of the compact and extended disks $\lesssim$ 10 au can be directly compared to the microlensing surveys (e.g., \citealt{Suzuki_2016b}.

In this paper, we aim to provide one of the first statistical results of substructures in compact disks by using the model-fitting approach employed in \citet{Zhang_2016} and \citet{Long_2020} to revisit the subsample of smooth, compact disks observed by \citet{Long_2019} in high-resolution ALMA imaging of 32 protoplanetary disks in the Taurus Molecular Cloud.  In Section \hyperref[sec:2]{2}, we briefly describe target selection and observations of the Taurus sample.  In Section \hyperref[sec:3]{3}, we outline our model-fitting approach, describe how substructures are characterized, and present the results of substructures identified in the subsample of compact disks that were previously classified as smooth. We compare the frequency, location and properties of substructures found in compact and extended disks within the full Taurus sample, and the ALMA DSHARP survey of nearby disks in Ophiuchus, Lupus and Upper Sco (see \citealt{Andrews_2018b}). In Section \hyperref[sec:4]{4}, we use the planet-disk interaction models of \citet{Zhang_2018} and simulated visibilities from a representative compact and extended disks to ascertain the sensitivity of our substructure detection technique to planets of various masses and separations. In Section \ref{sec:5}, we then calculate potential planet masses by gap widths using the method in \citet{Zhang_2018} and present them on the planet-mass-semi-major-axis diagrams. In Section \ref{sec:6}, we calculate occurrence rates and compare them to both potential planets in the DSHARP sample and the results of microlensing, direct imaging and radial velocity surveys.  We summarize our findings in Section \ref{sec:7}.

% SZ: We also calculate the potential planet masses and calculate the occurrences rates and compare them with the ones in the DSHARP sample and direct imaging, microlensing, and radial velocity surveys.

\section{Observations}
\label{sec:2}

Observations of the full 32 disk sample were conducted as part of ALMA Cycle 4 program 2016.1.01164.S (PI: Herczeg).  The goal of this observing program was to obtain a minimally-biased, high spatial resolution ($\thicksim$0.1\arcsec) sample of the full range of disk types around solar-mass stars in the Taurus star-forming region.  Disks around stars of spectral type later than M3 (to ensure sufficient S/N), known binaries with separations less than 0.5\arcsec, stars with high extinction (A$_V >$ 3 mag), and disks with existing high-resolution ALMA observations were excluded from target selection. The most significant bias of the sample selection comes from the avoidance of existing high resolution observations. Since the existing high resolution observations are biased towards bright and large disks, the Taurus sample can be biased against them. The selected targets sampled a considerable range of disk millimeter brightness, though the exclusion of close binaries naturally avoided some faint disks.  Additional details on target selection, and a table of host star properties, are provided in \citet{Long_2019}.

Continuum emission from the selected disks were observed in late August and early September of 2017.  Spectral windows at 218 and 233 GHz, both with identical 1.875 GHz bandwidths, were used.  Average observing frequency was 225.5 GHz, corresponding to a wavelength of 1.33 mm.  On-source integration times were between 4 and 10 minutes per target (see Table 2 of \citealt{Long_2019} for a complete observing log).  The C40-7 antenna configuration of ALMA Cycle 4, with baselines of 21 - 3697 meters (15 – 2780 k$\lambda$), was used to carry out the observations. 

The data were reduced using the Common Astronomy Software Applications (CASA) package (\citealt{McMullin_2007}), versions 4.7.2 and 5.1.1.  Following the standard ALMA pipeline, phase adjustments were made based on the water vapor radiometer measurements.  Bandpass, flux and gain calibrations were then applied for each measurement set.  From the calibrated visibilities, continuum images were created with the CASA task \textit{tclean} to perform phase and amplitude self-calibrations on targets with S/N $\gtrsim$ 30 (see \citealt{Long_2018} for greater detail).  Data visibilities were then extracted from the self-calibrated measurement sets and final continuum images were produced with Briggs weighting in \textit{tclean}.  The typical beam size of the final continuum images was 0.14\arcsec{} x 0.11\arcsec and typical RMS noise was 50 $\mu$Jy beam$^{-1}$.

\section{Modeling and Results}
\label{sec:3}

Of the 32 observed disks, 12 showed evidence of substructure following the approach of \citet{Long_2018}.  They determined the major axis of each disk by fitting an elliptical Gaussian profile to the continuum image with CASA task \textit{imfit}.  The radial intensity profile along the major axis was then inspected for evidence of substructure (e.g., inner cavities, extended emission at large radii, resolved rings or emission bumps) not able to be fit with a single smooth central component.  Model intensity profiles were produced by combining a central Gaussian profile (or exponentially tapered power law) with additional Gaussian rings inspired by peaks (if any) in the radial profiles.  Best-fit models (including the disk properties of position angle, inclination and phase center offsets) were obtained by comparing model and data visibilities, wherein model visibilities were created by a Fourier transform of the model intensity profile and matched to data visibilities with the Markov chain Monte Carlo method to compute the optimal value of free parameters.  Total flux and disk radius were then estimated from the best-fit models.

We adopt here the computed disk properties (see Table \ref{tab:3} in the Appendix), but use an alternative model-fitting approach directly in the visibility domain to reexamine the subsample of 12 smooth disks around single stars for evidence of small-scale substructure not visible from the continuum images.  We do not consider the 8 smooth Taurus disks in multiple star systems as they are not compatible with our fitting routine. The Taurus sample has 24 disks in single star systems. To produce a homogeneous sample for comparison between compact and extended disks (see the definition in Section \ref{sec:comparecompactextended}), we also apply our fitting routine on 12 disks with substructures already found in \citet{Long_2018}.

\subsection{Model-fitting Approach}
\label{sec:3.1}

In the model-fitting approach employed here (see \citealt{Pearson_1999} for a full treatment), data visibilities are reproduced via a parametric model of the source intensity distribution.  Crucially, radial intensity profiles are created by fitting directly in the visibility domain.  In the original approach (\citealt{Long_2019}), intensity profiles are retrieved from the continuum images and visibility fitting is used to optimize model profiles.  By starting in the visibility domain, we utilize the full complement of spatial frequency information within the data and can recover smaller scale substructure than can be seen directly in CLEANed continuum images (\citealt{Zhang_2016}).

Visibilities are deprojected using the inclination and position angles given in Table \ref{tab:3}.  For circularly symmetric disk emission, the deprojected visibilities and radial brightness distribution are related by a Hankel transform (\citealt{Pearson_1999}): 

\vspace*{5pt}

\begin{equation}
u' = (u \cos \phi - v \sin \phi) \times \cos i
\end{equation}

\begin{equation}
v' = u \sin \phi + v \cos \phi
\end{equation}

\begin{equation}
V(\rho) = 2\pi \int_{0}^{\infty} I_{\nu}(\theta)\theta J_{0}(2\pi\rho\theta)d\theta
\end{equation}

\vspace*{10pt}

\noindent where $i$ and $\phi$ are the disk inclination and position angle, $\rho = \sqrt{u'^{2}+v'^{2}}$ is the deprojected uv-distance in units of $\lambda$, $\theta$ is the radial angular scale from the center of the disk, and $J_{0}$ is a Bessel function of the first kind.

We model $I(\theta)$, the disk intensity distribution, with a parametric function (Equation \ref{eq:4}) developed by \citet{Zhang_2016}.  This parametric function is characterized by a series of Gaussian functions, each modulated by a sinusoidal function with a spatial frequency of $\rho_{i}$.  The number of Gaussian functions is determined by the number of distinctive peaks in the disk visibility profile.  A peak in visibility indicates that some particular spatial frequency contributes more than others.  Variables \{$a_{0}, \sigma_{0}, a_{i}, \sigma_{i}, \rho_{i}$\} are free parameters.  As such, 

\begin{equation}
\label{eq:4}
\begin{aligned}
I(\theta) = {} & \frac{a_{0}}{\sqrt{2\pi}\sigma_{0}}\exp\left(-\frac{\theta^{2}}{2\sigma_{0}^{2}}\right) \\ & + \sum_{i} \cos(2\pi\theta\rho_{i}) \times \frac{a_{i}}{\sqrt{2\pi}\sigma_{i}}\exp\left(-\frac{\theta^{2}}{2\sigma_{i}^{2}}\right).
\end{aligned}
\end{equation}

\begin{figure*}[t!]
\centering
\vspace{-3.5cm}
\includegraphics[width=0.99\textwidth]{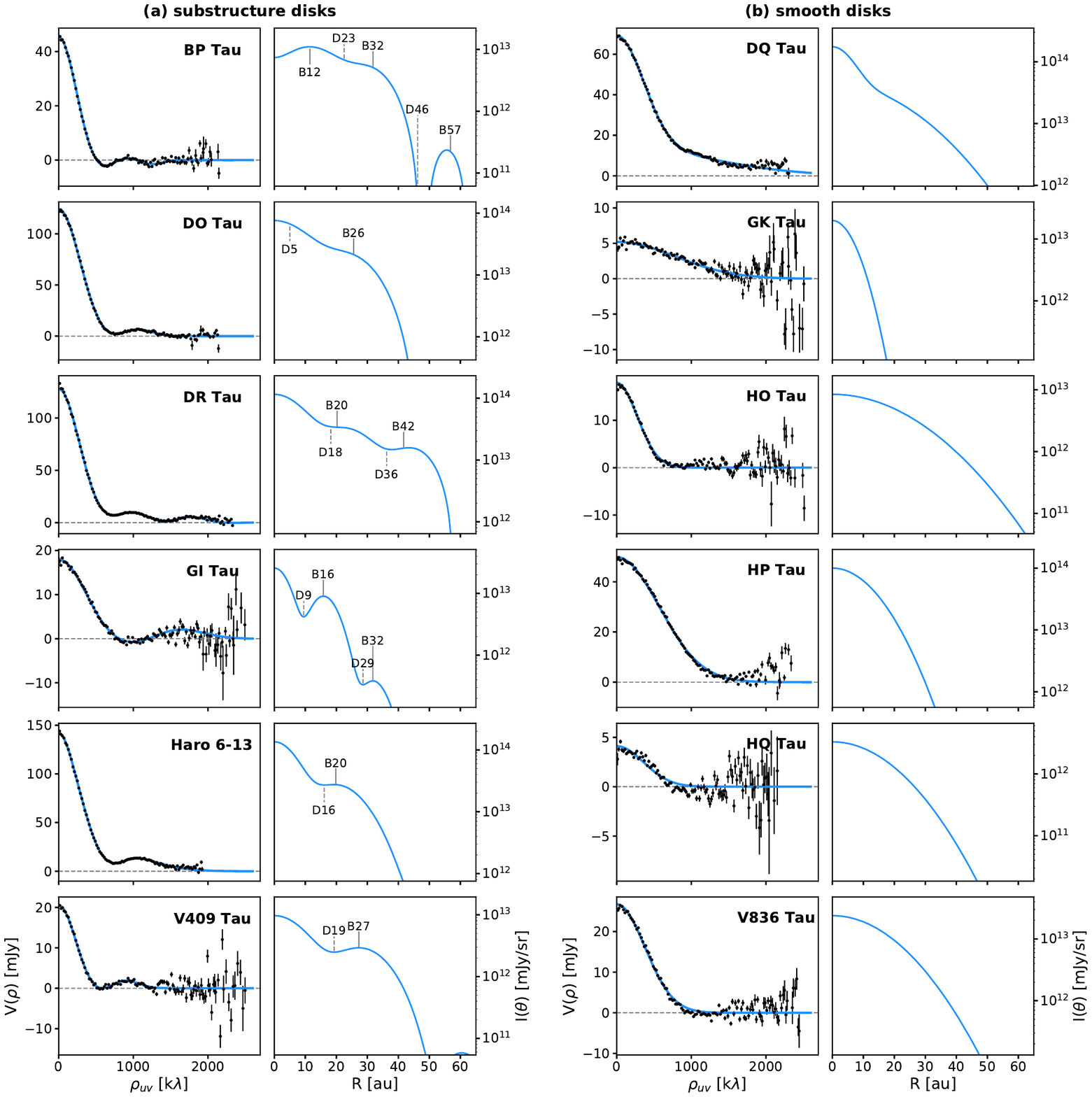}
\vspace{-4cm}
\caption{Deprojected visibility and radial intensity profiles for the six disks with identified substructure (a) and the six smooth disks (b) following our fitting approach.  Overlaid in blue on the visibility curves are our best-fit models, which are used to derive the adjacent radial intensity profiles.  Dashed black lines on the radial intensity curves of panel (a) mark gaps, and solid gray lines mark rings.}
\label{fig:1}
\end{figure*}

We perform model-fitting with the MPFIT routine (\citealt{Markwardt_2009}).  This routine iteratively searches for optimal values of the free parameters using the Levenberg-Marquardt technique.  We provide initial guesses for the amplitudes \{$a_{i}$\}, widths \{$\sigma_{i}$\} and central locations \{$\rho_{i}$\} of the identified peaks in each visibility profile.  The central location of the first peak is invariably set to zero, and we avoid attempting to fit additional Gaussians to the noisy region seen at baselines beyond $\sim$1650 k$\lambda$. For GK Tau, HO Tau, HP Tau, HQ Tau, and V836 Tau, the visibilities are very noisy beyond $\sim$1000 k$\lambda$, and no clear bump can be identified to justify additional Gaussian components, so only one Gaussian components are used. Additionally, the number of Gaussians used in a fitting can be justified by the chi-square values. In Appendix Figure \ref{fig:14}, we demonstrate that two-Gaussian component models of DO Tau and DQ Tau reduce the chi-square values by an order of magnitude compared to that of one-Gaussian models. The best-fit model visibilities for the 12 compact disks in our sample are shown in columns one and three of Figure \ref{fig:1}.
The model-derived radial intensity profiles are shown in columns two and four. The best-fit parameters and chi-square values are listed in Appendix Table \ref{tab:1p}. The reanalyses of 12 disks in \citet{Long_2018} are shown in Appendix Figure \ref{fig:13} and Table \ref{tab:5a}. We quantify and report the labeled features of the radial intensity profiles in the next two subsections.

\subsection{Substructure Characterization}
\label{sec:3.2}

We identify substructures in our model radial intensity profiles by searching for local minima and maxima along the curve.  Local minima, or dips, in the curve are interpreted as gaps and labeled with the prefix D.  Local maxima, or bumps, are interpreted as rings and labeled with the prefix B.  The number following each prefix identifies the radial location of the gap or ring (e.g., B12 refers to a ring identified at or near 12 au).  This convention follows from the literature (e.g., \citealt{ALMA_2015}; \citealt{Huang_2018}).

Shallow features in the radial intensity profiles not immediately identifiable as gaps or rings, but which deviate from a smooth profile, are characterized as ``plateaus" according to the criteria established in \citet{Huang_2018}.  Inner and outer edges are identified by visual inspection and the deviation is deemed worthy of inclusion if $\frac{1}{I_{v}(r)}\frac{dI_{v}(r)}{dr}$, the slope of the radial intensity divided by the radial intensity, exceeds -0.05 (i.e., if the radial decrease of the intensity is small).  The inner and outer edges of plateau features are also labeled with the prefixes D and B, respectively.

We measure the widths and depths of gaps and rings following the approach of \citet{Huang_2018}.  The gap outer edge, and the ring inner edge, are defined as the radius at which model intensities most closely match $I_{mean} = 0.5(I_{d} + I_{b})$.  The gap inner edge is the radius interior to gap center with an intensity value closest to $I_{mean}$, and the ring outer edge the radius exterior to ring center with an intensity value closest to $I_{mean}$.  Widths are then a simple subtraction of inner and outer edges.  Gap depths are estimated by $\frac{I_{d}}{I_{b}}$, a ratio of the intensities at gap and ring centers.  Widths of the plateau features are measured singly as we identify not gaps and rings, but an inner and outer plateau edge.  Plateau depths are estimated by fitting a spline function to data on either side of the deviation and measuring how much it varies from the fit.

\subsection{Model-fitting Results}

Substructures are identified in 6 of the 12 compact disks (see left-hand columns of Figure \ref{fig:1}), and labeled following the convention outlined in the previous subsection.  The properties of these substructures are listed in Table \ref{tab:1}.  We note similarities to the recent results of \citet{Jennings_2022b}.\footnote{They use the non-parametric Frankenstein code (\citealt{Jennings_2020}) to directly fit the visibilities from all 24 Taurus disks in single star systems.  Of the subsample of 12 compact disks that we reanalyze in this paper, they claim substructure detections in BP Tau and DR Tau consistent with what we identify here.  They also note tentative signs of substructure in DO Tau and V409 Tau.  We report different results for GI Tau and Haro 6-13, but this could be due to an overaggressive fit of the long baseline data from GI Tau and the shallowness of the feature in Haro 6-13.  In general, their published visibility and radial profiles are similar to what we produce here, especially at the shorter baselines.}  All 6 disks with substructure show at least one gap-ring pair (or plateau) interior to 45 au.  BP Tau and DO Tau show evidence of plateau features between 23 and 32 au and 5 and 26 au, respectively.  The midpoints of these features (27 and 15 au) could correspond to gap center in future high-resolution data.  No gaps are detected interior to 9 au, but the lack of emission in the core of BP Tau hints at substructure.  In general, the dearth of structure in the inner disks might be expected given the resolution of our observations ($\sim$15 au at the typical distance to the Taurus region) and the high optical thickness of many disk cores (\citealt{Ansdell_2018}; \citealt{Huang_2018}).  In the case of GI Tau, we note that the bump in its visibility profile occurs in the relatively noisy region beyond 1500 k$\lambda$ and the two reported features may be an artifact of an overaggressive fit.  It is otherwise one of the faintest disks in our 12-disk sample.  Feature locations tend to cluster near the middle of disk continuum emission, but this could be a consequence of the aforementioned resolution and optical depth constraints at disk center, and the faintness of disk emission at extended radii.  No preferred radial location for substructure features is otherwise obvious.

\begin{table}[t!]
\centering
\textbf{Table 1} \\
\text{Characteristics of Identified Substructures} \\
\smallskip
 \begin{tabular*}{0.4725\textwidth}{c @{\extracolsep{\fill}} cccc}
 \hline \hline
 Disk & Feature & $r_0$ (au) & Width (au) & Depth ($\frac{I_{d}}{I_{b}}$) \\
 \hline
 BP Tau & B12 & 11.5 & 11.4 & - \\
        & D23 & 22.5 & 9.33 & 0.985 \\
        & B32 & 31.8 & - & - \\
        & D46 & 45.9 & 8.17 & $\star$ \\
        & B57 & 57.1 & 8.38 & - \\
 \hline
 DO Tau & D5 & 5.00 & 20.6 & 0.853 \\
        & B26 & 25.6 & - & - \\
 \hline
 DR Tau & D18 & 18.2 & 1.71 & 0.996 \\
        & B20 & 20.2 & 1.83 & - \\
        & D36 & 36.2 & 4.71 & 0.932 \\
        & B42 & 41.7 & 5.05 & - \\
 \hline
 GI Tau & D9 & 9.47 & 5.31 & 0.462 \\
        & B16 & 15.8 & 6.75 & - \\
        & D29 & 28.6 & 2.31 & 0.875 \\
        & B32 & 31.7 & 3.22 & - \\
 \hline
 Haro 6-13 & D16 & 16.1 & 3.06 & 0.965 \\
           & B20 & 19.9 & 3.58 & - \\
 \hline
 V409 Tau & D19 & 19.3 & 6.59 & 0.850 \\
          & B27 & 27.2 & 7.86 & - \\
 \hline
 \vspace{-2.5mm}
 \end{tabular*}
{\raggedright \textbf{Notes:} $r_0$ represents the radial distance from disk center.  No depth measurement is provided for D46 of BP Tau as the intensity at gap center is suspiciously low and the gap-ring pair itself may be an artifact of fitting noisy data at long baselines (see \citealt{Jennings_2022b} for a discussion). \par}
\rtask{tab:1}
\end{table}

With the exception of the standalone ring at 12 au in BP Tau, all non-plateau features have widths of 1 to 8 au (see Table \ref{tab:1}).  There is no clear correlation between feature location and width, but all features outside of 15 au (equivalent to the resolution limit) have width to radial location ratios between 0.08 and 0.43.  This is roughly consistent with what has been observed in other samples (e.g., \citealt{Huang_2018}).  All but a single gap have listed depths greater than or equal to 0.85 (i.e., intensity contrasts between adjacent gaps and rings of 15\% or less).  On average, these are shallower than the gaps reported in \citet{Long_2018} and \citet{Huang_2018}, which could explain why they were overlooked.  As implicated before, the gap-ring pair of GI Tau which shows an intensity contrast in excess of 15\% may be the result of a tenuous fit.  The wide (and noisy) bump at long baselines within its deprojected visibility profile could instead be fit with a sharp power law (see Figure 11 of \citealt{Long_2019}).

No robust features are detected in the other six disks of the sample (see right-hand columns of Figure \ref{fig:1}).  While these smooth disks are on average both fainter and more compact than those with substructure, they do begin to occupy the same parameter space, which begs the question of what else could separate these two populations.  Any hidden substructure, especially in the outer disks, might be hard to identify given their meager total disk fluxes (see \citealt{Long_2019}).  It's notable that the brightest of these six smooth disks, DQ Tau, has a deviation from a smooth curve visible around 15 au, but this deviation does not meet the -0.05 slope criteria established previously for plateau features.

In Appendix Figure \ref{fig:13}, we also present the radial profiles of 12 disks with previously identified substructures in \citet{Long_2018} using our fitting routine. Since these disks generally have more features in the visibility planes, their radial profiles can have very deep gaps below the sensitivity limit of any state-of-the-art observations. Thus, we make a cutoff at 10$^{12}$ mJy/Sr, the sensitivity limit for the DSHARP observations \citep{Andrews_2018b}. Emissions below this value is taken as this cutoff value. Their corresponding substructures are listed in Table \ref{tab:5a} in the Appendix. Overall, we find these profiles are similar to those in \citet{Long_2018}, but with more substructures in the inner disks. With our routine, some of the gaps at the outer disks disappear due to the weak emissions. More details can be found in the Appendix.

\subsection{Comparison of Compact and Extended Disks \label{sec:comparecompactextended}}

We compare our results of the 24 Taurus disks (\citealt{Long_2018}) and the 18 disks in the DSHARP sample (\citealt{Andrews_2018b}; \citealt{Huang_2018}).  We do not include substructures identified in the relatively small number of additional disks observed at (generally) lower spatial resolutions (see Table 5 of \citealt{Huang_2018} for a partial listing).

We classify disks with $R\textsubscript{eff,90\%}$ (the radius which encloses 90\% of total disk continuum flux) less than 50 au as compact and disks with $R\textsubscript{eff,90\%}$ greater than or equal to 50 au as extended.  This cutoff is motivated by the finding in \citet{Long_2019} that all Taurus disks with $R\textsubscript{eff,95\%} \ge$ 55 au show detectable substructure, and similar cutoff choices in the literature (e.g., \citealt{van_der_Marel_2021, Jennings_2022b}).  The average effective radius of our 14 (2 in \citealt{Long_2018}, and 12 in \citealt{Long_2019}) compact disks is $\sim$ 33 au.

Substructures are detected in 12 of 18 compact disks, and all 24 extended disks (following the results of \citealt{Long_2018} and \citealt{Huang_2018}), in single star systems.\footnote{Smooth compact disks: \textit{HO Tau, HP Tau, DQ Tau, V836 Tau, GK Tau and HQ Tau}  \\ \\Compact disks with substructure: \textit{WSB 52, DoAr 33, SR 4, HD 142666, FT Tau, IP Tau, DO Tau, V409 Tau, DR Tau, Haro 6-13, BP Tau and GI Tau} \\ \\Extended disks: \textit{GW Lup, DoAr 25, Sz 114, IM Lup, Sz 129, HD 143006, Elias 24, RU Lup, WaOph 6, AS 209, HD 163296, MY Lup, Elias 20, Elias 27, UZ Tau, DS Tau, MWC 480, RY Tau, GO Tau, IQ Tau, DN Tau, CI Tau, DL Tau and CIDA 9}} Our detection of substructure in six compact Taurus disks at least doubles the number of compact disks with claimed detections in the literature (e.g., \citealt{Gonzalez-Ruilova_2020, Kurtovic_2021}), and allows for a more robust initial comparison of substructure in small and large disks.

\begin{figure}[t!]
\centering
\includegraphics[width=0.495\textwidth]{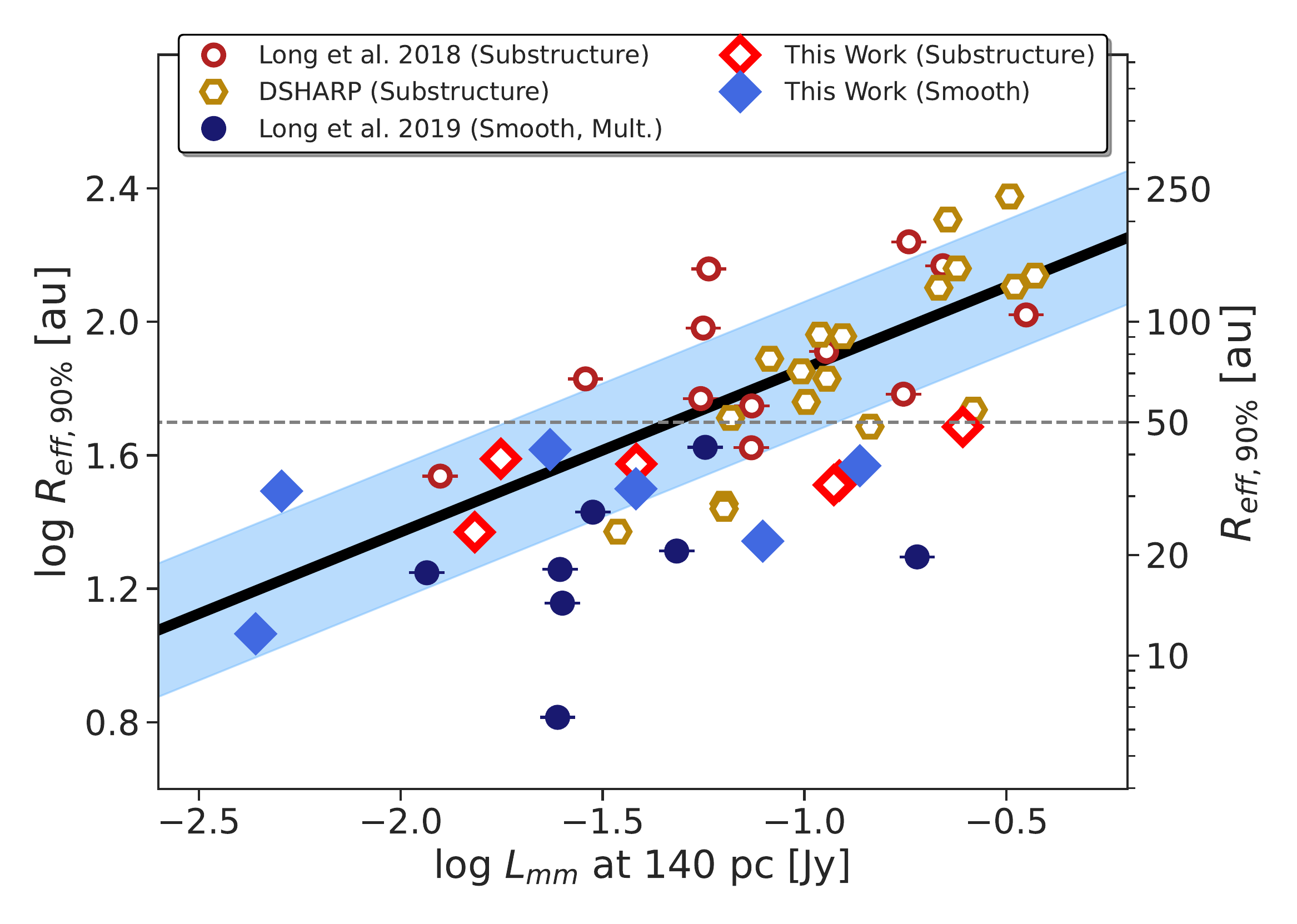}
\caption{Disk continuum luminosity-size relationship for the DSHARP and full Taurus samples.  The dark blue circles represent the subsample of eight smooth disks in multiple star systems from \citet{Long_2019}.  Our chosen cutoff of 50 au effective radius for compact and extended sources is marked by a dashed gray line.  The solid black line denotes the millimeter scaling relation (scaled to $R\textsubscript{eff,90\%}$) observed by \citet{Andrews_2018a} in their sample of 105 nearby protoplanetary disks.  The light blue shading represents the 68\% confidence interval (plus an additional scatter term) of that relation.}
\label{fig:4}
\end{figure}

We present the disk luminosity-radius relationships for the DSHARP and full Taurus samples in Figure \ref{fig:4}.  We scale luminosities as $F_{\nu}(d/140)^2$ to a standard distance of 140 pc, with $F_{\nu}$ denoting the total flux from the disk at $\sim$1.3 mm.  Scaled luminosities and effective radii from the samples generally follow the scaling relationship observed by \citet{Andrews_2018a} in their large sample of nearby protoplanetary disks, with the caveat that those observations were taken at a wavelength of 870 $\mu$m.  Our six disks with identified substructure begin to occupy the same parameter space probed by \citet{Long_2020} in their analysis of the compact GQ Lup disk.  While our six disks without identified substructure are on average $\sim$50\% fainter and $\sim$30\% more compact than their non-smooth counterparts, three of them (DQ Tau, HO Tau and V836 Tau) occupy positions on the luminosity-size plot where substructures have been detected in similar disks.  The discrepancy then is mainly driven by just three extra faint and/or compact disks: HP Tau, GK Tau and HQ Tau.  We again note that the emission profile of DQ Tau (as seen in Figure \ref{fig:1}) does present a deviation from a smooth profile that was very nearly classified as a plateau feature following the classification of Section \ref{sec:3.2}.  Additional disk substructures very likely exist, but do not lend themselves to easy detection because they are too narrow to be resolved, or are present in the extreme inner or outer disk.

We do not apply our model-fitting approach to the sample of eight smooth disks in multiple star systems from \citet{Long_2019}, but their sizes and luminosities broadly overlap those of the six smooth disks in single star systems.  No substructures are detected in disks with $R\textsubscript{eff,90\%}$ less than 23 au or continuum luminosities below 12\,mJy.  This underlines the need for better spatial resolution observations of the faintest and most compact disks, with the caveat that those disks may simply be void of substructure.

\begin{figure*}[t!]
\centering
\includegraphics[width=0.99\textwidth]{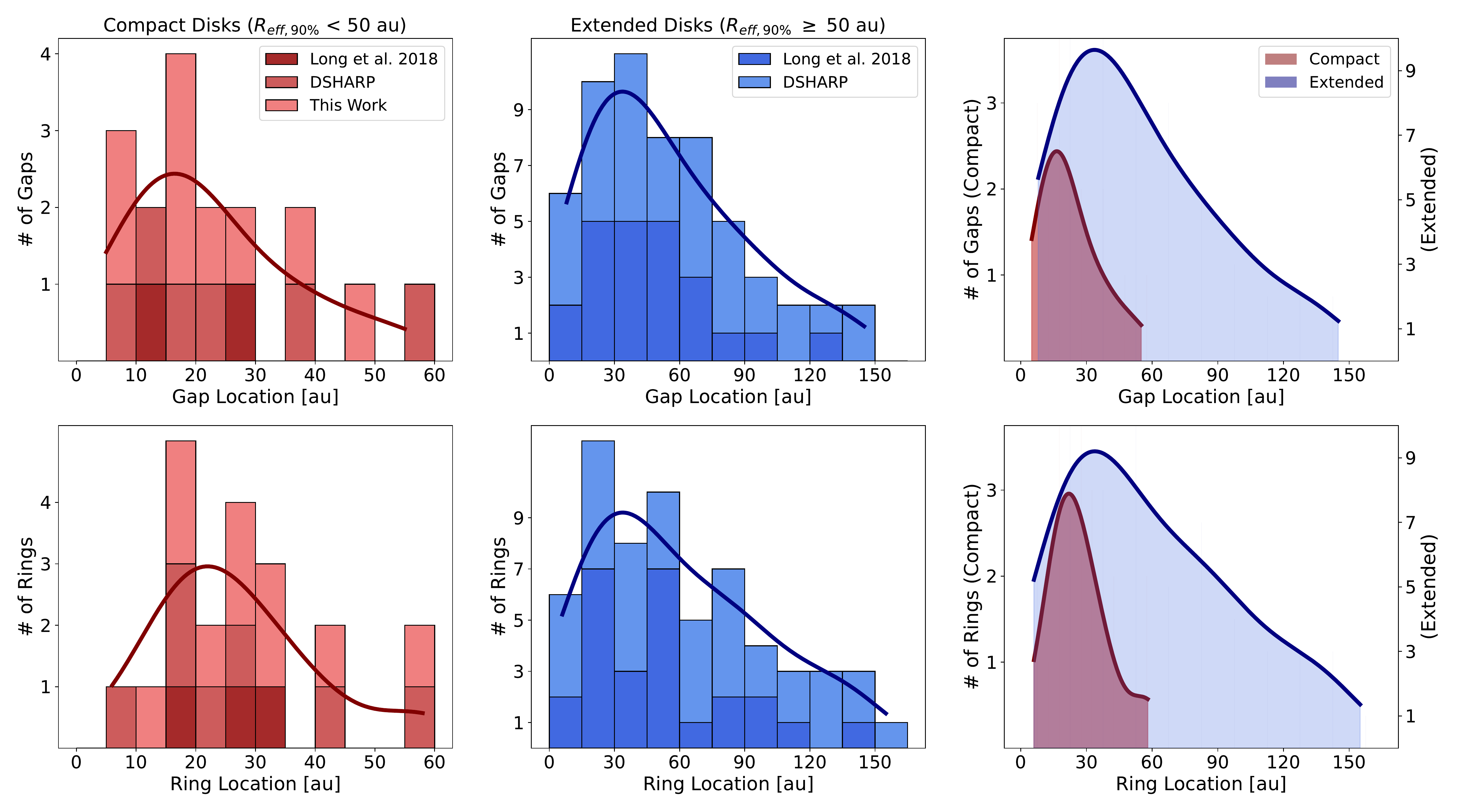}
\caption{Incidence of gaps and rings in 5 au radial location bins for the compact case (left-hand panels), and 15 au bins for the extended case (middle panels).  Probability density functions (PDFs) of substructure locations (scaled to match the underlying histograms) are overlaid in red and blue, respectively.  These PDFs are plotted together in the right-hand panels for a more direct comparison of gap and ring locations in compact and extended disks.}
\label{fig:5}
\end{figure*}

We display the incidence and location of gaps and rings among the compact and extended samples in the histograms of Figure \ref{fig:5}. The locations of the gaps and rings in 12 extended disks studied by \citet{Long_2018} are updated with results of our visibility fitting (listed Table \ref{tab:5a}). Gap locations peak between 15 and 20 au in the compact case and taper off out to $\sim$55 au.  The detection of gaps (and rings) past 50 au is possible because emission extends beyond the effective radius threshold of 90\% that we use in our determination of a compact disk.  
Ring locations peak near 20 au and follow much the same trajectory as the gaps.

Considering the extended case, gap locations peak near 40 au and exhibit a slow decline out to $\sim$145 au.  Ring incidence peaks near 30 au, and declines in a similar trend.  In general, substructure is identified out to large relative separations in both compact and extended disks and does not strongly cluster around any one location.  Furthermore, what we describe as a peak is likely an exaggeration of the true distribution given the increased difficulty of detecting features near disk center (high optical depths, smaller characteristic sizes) and in the faint outer regions, and also the rarity of very large disks.  The apparent decline is also a function of plotting so many disks together.  Some disks have smaller effective radii than others in the same category (compact or extended) and cannot therefore have substructure at the most distant separations plotted (\citealt{Huang_2018}).  This is highlighted by the fact that every disk with identified substructure in our sample has at least one feature near, at or beyond its calculated effective radius.  Such a finding is perhaps not surprising if we invoke substructure as a necessary condition to stop the fast radial drift of disk solids predicted under the assumption of a smooth gas disk (e.g., \citealt{Whipple_1972}; \citealt{Weidenschilling_1997}).  Ultimately, substructure may be a common occurrence at all radii in both compact and extended disks.

\begin{figure}[t!]
\centering
\includegraphics[width=0.495\textwidth]{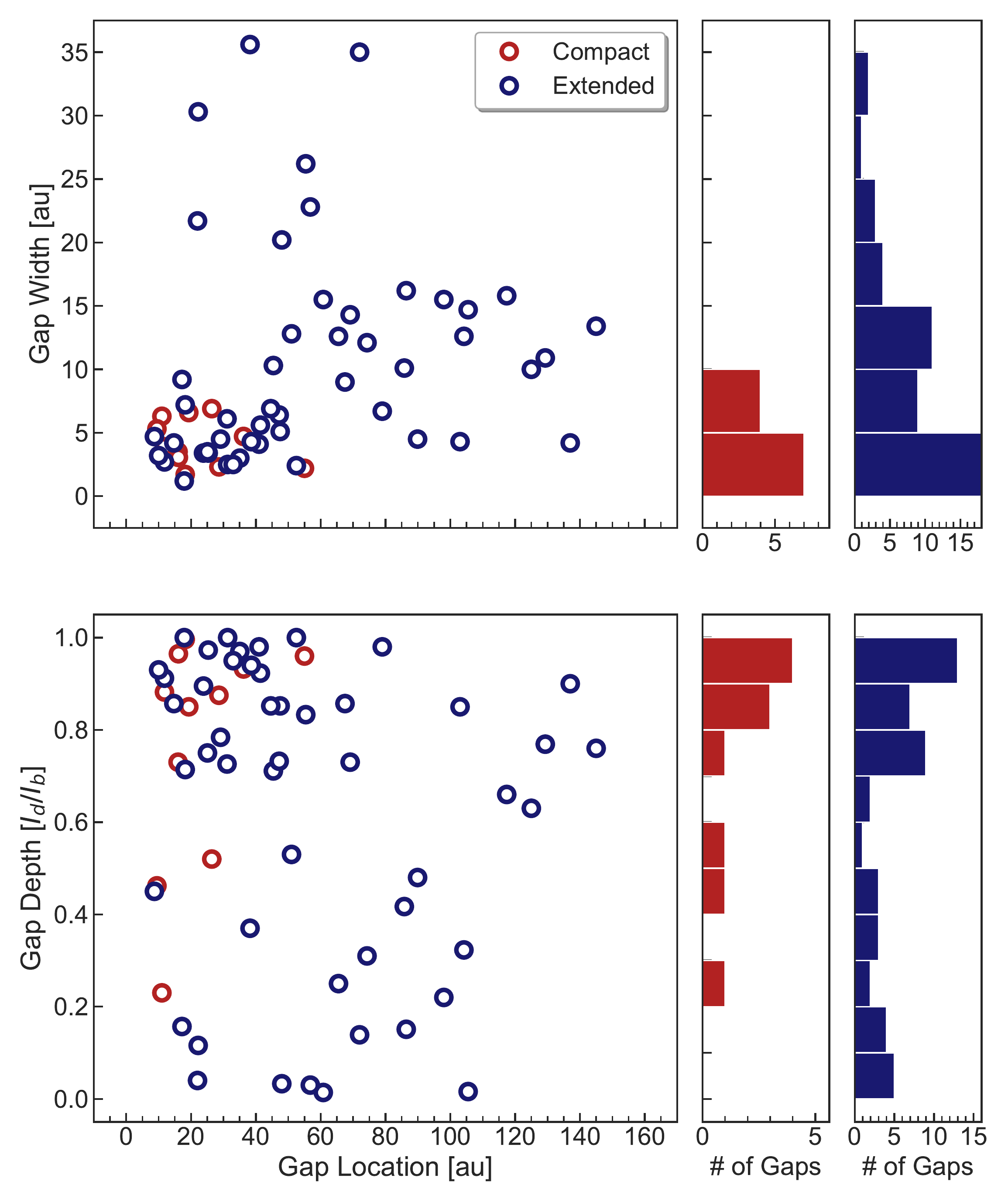}
\caption{Comparison of gap widths (top panels) and gap depths (bottom panels) for gap-ring pairs in the compact and extended disks of \citet{Long_2018}, \citet{Huang_2018} and this paper.  Gap depths are as described in Section \ref{sec:3.2}, with values near unity representing shallow features and values near zero representing deep features.}
\label{fig:6}
\end{figure}

We compare gap widths and depths for compact and extended disks in the combined plots of Figure \ref{fig:6}.  We exclude the four inner disk cavities of \citet{Long_2018} without defined gap locations, and all plateau features identified in the DSHARP sample and this paper.  The widths and depths of these plateau features are either not calculated, or calculated differently than the other features. This leaves 11 gaps in the compact disks and 48 gaps in the extended disks for analysis.

All the gaps in compact disks span widths of less than 10 au, values near the resolution limits of the observations.  \citet{Huang_2018} note that beam effects near the resolution limit will generally deflate the measurement of gap widths, and we might therefore expect that our reported widths are an underestimate.  Still, gap widths much larger than 10 au would begin to exceed the effective radii of the smallest disks and would be a counterintuitive result in a compact disk.  Gap widths in the extended case are more varied, with several exceeding 20 au, but the majority are smaller than 15 au.  In both cases, narrow widths cluster in the inner 50 to 60 au.  No obvious correlation of gap width and location otherwise exists.

\begin{figure}[t!]
\centering
\includegraphics[width=0.495\textwidth]{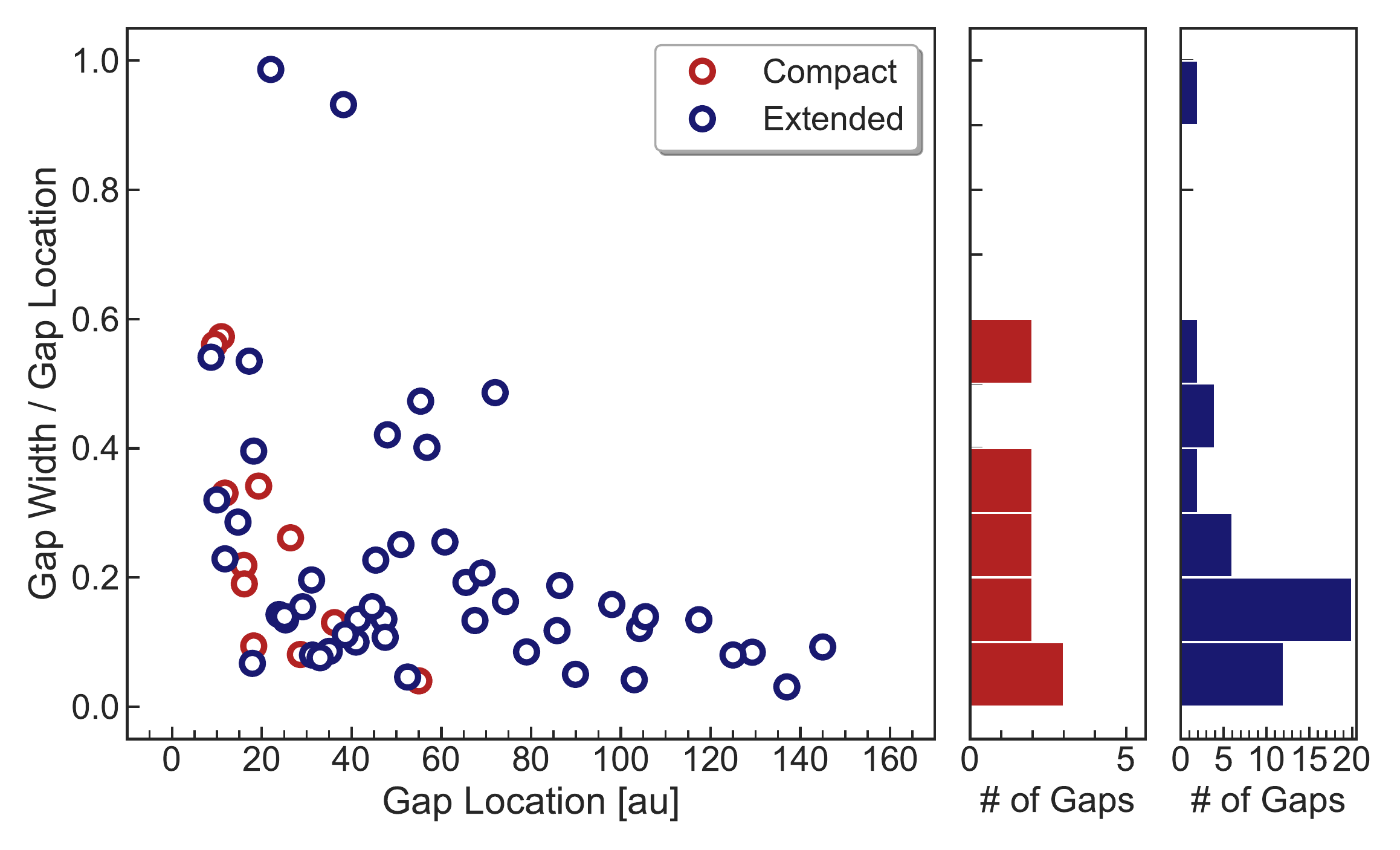}
\caption{Normalization of gap width to gap location for compact and extended disks.  This ratio is a better indicator of potential planet mass than absolute gap width and depth alone.}
\label{fig:7}
\end{figure}

The story is much the same with gap depths.  The majority of gaps are shallow, with gap-ring intensity variations below 30\%, especially interior to 60 au.  It is tempting to link wider and deeper gaps at large radii in the extended disks to a more massive planet population (more massive planets carve out larger gaps under the assumption of planet-disk interaction), but this relationship largely disappears after correcting gap width to gap location (Figure \ref{fig:7}), a better indicator of planet mass \citep{Kanagawa_2016, Zhang_2018}. It is worth mentioning that gap widths are on average smaller when the gaps are between 20-50 au, except two very large ones. Wide gaps are more clustered between 10-20 au and 50-80 au. Since planet masses are inferred from the gap width, the paucity of giant planets between 30-50 au can be found in all following planet population figures.  We extend this discussion in Section \ref{sec:5.2}. We also note that the uncertainties are larger in the inner disk where low-mass planets with shallow gaps are especially difficult to study.

\section{Planet-Disk Interactions: \\Detection Limit}
\label{sec:4}

Planet-disk interactions are proposed as one of the major causes of observed disk substructure (e.g. \citealt{Rice_2006}; \citealt{Zhu_2012}), and we expect that gap location should coincide with orbital location of the embedded planet.  Additionally, more massive planets will form steeper pressure gradients within the disk and create larger gap openings than are created by less massive planets (\citealt{Fung_2014}; \citealt{Kanagawa_2015}; \citealt{Rosotti_2016}; \citealt{Zhang_2018}).

In order to characterize the sensitivity of our model-fitting approach for detecting substructure induced by young planets, we combine a series of planet simulations with a sample compact disk.

\subsection{Use of Planet-Disk Interaction Models}

We adopt dust surface brightness distributions from the planet-disk interaction models of \citet{Zhang_2018}.  These 2D simulations include both gas and dust components, and were originally used to explore possible planet properties inferred from substructures identified in the sample of DSHARP disks (see \citealt{Huang_2018}) under the planet-disk interaction hypothesis.  At the radius of the planet, we select a disk aspect ratio ($h/r$) of 0.05, the $a_{\mathrm{max}}$ = 0.1 mm (DSD1) dust size distribution, and a disk surface density of 30 g\,cm\textsuperscript{-2} (see Section 2 of \citealt{Zhang_2018} for more detail on these parameters).  We vary planet mass, distance to host star and the disk viscosity condition (specifics to follow).  Total disk continuum flux at 1.33 mm is set at 50 mJy, a medium value in our sample of Taurus disks.  The mass of the host star is fixed at 1 $M_\odot$. The disks are assumed to be face-on.

To produce simulated visibilities, the 2D model FITS images are used as inputs to the Python package \textit{vis\_sample}\footnote{\url{https://github.com/AstroChem/vis_sample}} for fast Fourier transform and resampling of the visibility points at the same uv coordinates as the observations of the Haro 6-13 disk.  We choose Haro 6-13 to generate our models on the basis of its baseline coverage being typical of our Taurus sample.  The simulated visibilities are then deprojected and analyzed as per the same model-fitting approach outlined in Section \hyperref[sec:3.1]{3.1}.

We generate 90 model disks by varying planet mass between five values (11 $M_\earth$, 33 $M_\earth$, 0.35 $M_J$, 1 $M_J$ and 3.5 $M_J$), planet distance to host star from 5 to 30 au in increments of 5 au, and the disk viscosity condition ($\alpha$) between $10^{-4}$, $10^{-3}$ and $10^{-2}$.  More information 
%on the range of planet masses 
can be found in \citet{Zhang_2018}.  We vary the planet-star distance between 5 and 30 au in consideration of where ice and gas giant planets lie within our own solar system, and resolution constraints of our disk observations. We also place a 11 $M_\earth$ planet at 50 au, as it is a typical radius for potential planets in extended disks (shown in Figure \ref{fig:5}).  We use different values of the disk viscosity condition because this value is not well constrained by existing observations of disk phenomena (e.g., \citealt{Pinte_2016}; \citealt{Rafikov_2017}; \citealt{Flaherty_2018}; \citealt{Trapman_2020}), especially for compact disks.  We do not vary the disk aspect ratio, but note that higher values will open gaps that are shallower and wider than in the 0.05 case (\citealt{Zhang_2018}).  Our goal is to simulate ``average" conditions at the radii of our embedded planets.

\begin{figure}[t!]
\centering
\includegraphics[width=0.495\textwidth]{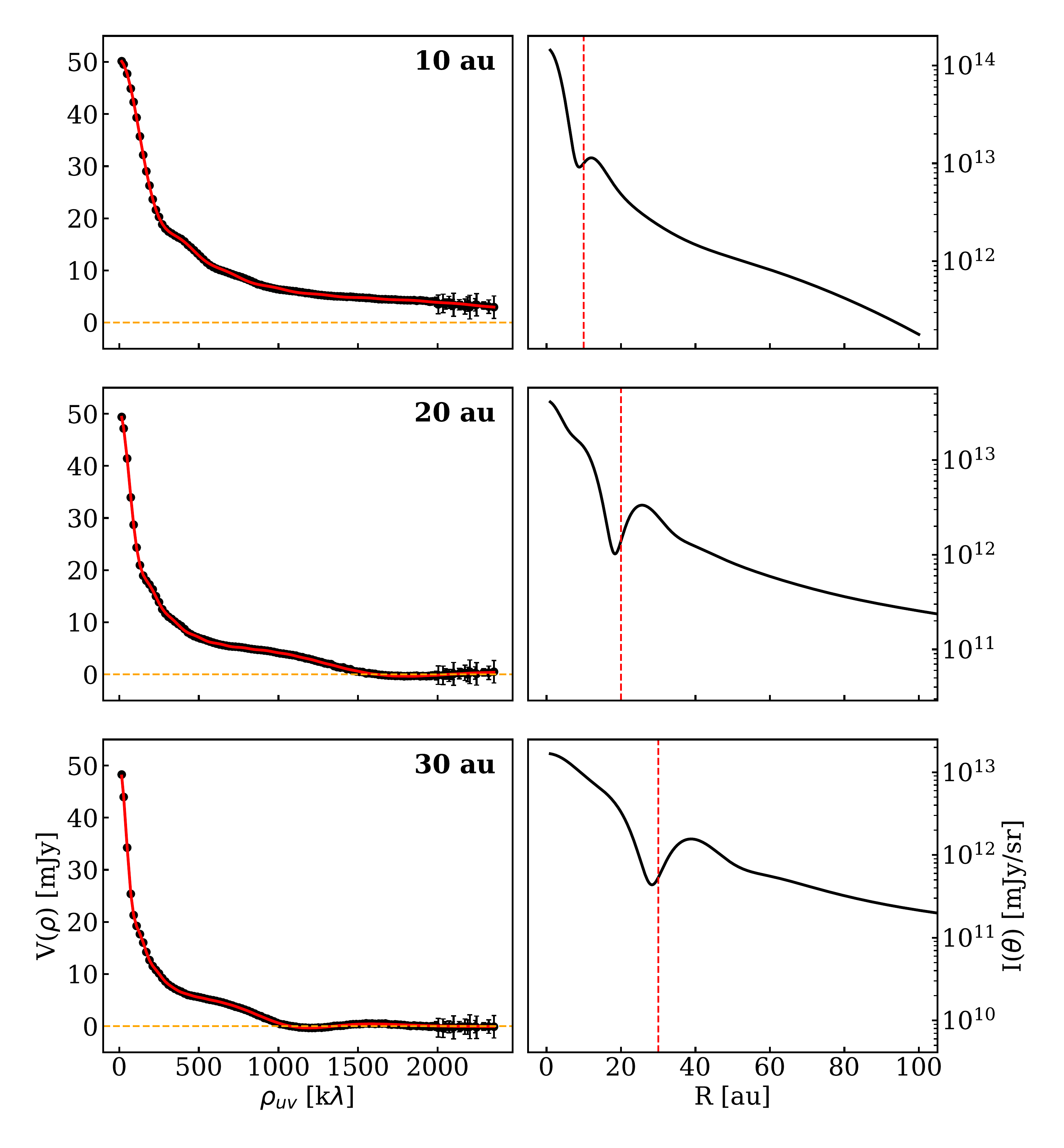}
\caption{Deprojected visibility and radial intensity profiles for an approximately Saturn-mass ($M_p$ $\approx$ 1.2 $M_{Saturn}$) planet injected at 10, 20 and 30 au under the $\alpha$ = $10^{-3}$ medium viscosity regime.  Our best-fit model of the visibility profile is overlaid in red in the left-hand column, and the location of each inserted planet is marked by a red dashed line on the adjacent radial intensity curves.  The orange dashed lines represent the zero point of the visibilities.}
\label{fig:2}
\end{figure}

Deprojected visibility profiles for three planet-star distances of the $M_p$ = 0.35 $M_J$ and $\alpha$ = $10^{-3}$ case are shown in the left-hand column of Figure \ref{fig:2}.  From the deprojected profiles, we produce a best-fit visibility model and use it to derive a radial intensity profile for each disk.  Best-fit models for these three median cases are produced with multiple Gaussians initiating the MPFIT routine.  As planet mass increases, the tendency is for more and larger bumps in the visibility profile, necessitating more Gaussians to initiate the fitting routine.

\begin{figure*}[t!]
\centering
\includegraphics[width=0.99\textwidth]{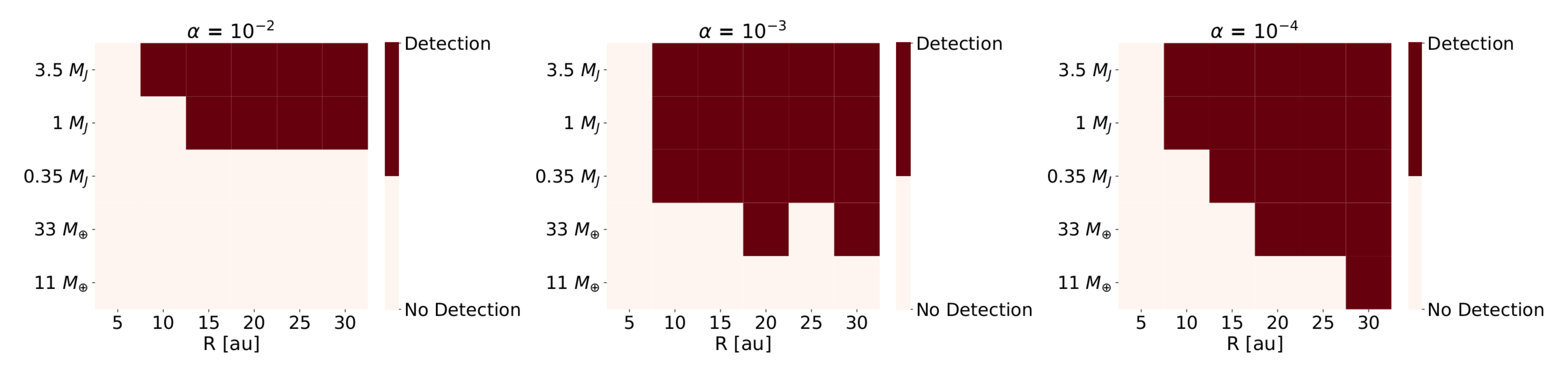}
\caption{Detections and non-detections of simulated planets from 5 - 30 au under three different disk viscosity conditions: $\alpha$ = $10^{-2}$ (left), $\alpha$ = $10^{-3}$ (middle) and $\alpha$ = $10^{-4}$ (right).}
\label{fig:3}
\end{figure*}

In order to classify gap-ring pairs near the locations of our inserted planets as detections or not, we borrow the $\frac{1}{I_{v}(r)}\frac{dI_{v}(r)}{dr}$ test employed previously to identify plateau features.  However, we now demand that the slope exceeds 0 instead of -0.05 (i.e., there must be a radial increase of intensity within the bounds of the gap-ring pair), since we want to focus on gaps and rings, but not plateaus. Gaps and rings are easier to infer planet masses from.

\subsection{Results from the Simulations}

We show the results from the model disks for each viscosity condition in the heat maps of Figure \ref{fig:3}. We assume planets within the detection zone have 100\% percent possibility of detection and 0\% out of the detection zone. For the $\alpha$ = $10^{-2}$ case, our model-fitting approach detects Jupiter-mass planets at separations of 15 au or greater from their host stars.  Sensitivity improves to 10 au for a multiple Jupiter-mass planet.  This mass-radius regime lies at the lower end of what is currently detectable with direct imaging techniques (e.g., \citealt{Macintosh_2015}; \citealt{Nielsen_2019}), and the upper end of what is detectable with the radial velocity technique.\footnote{See the \href{https://exoplanetarchive.ipac.caltech.edu}{NASA Exoplanet Archive} for an up-to-date mass-period plot.  We note that direct imaging techniques are insensitive to planets close to their host stars because of insufficient contrast.  Meanwhile, planets at far separations do not create sufficient wobble in their host stars to be detected by the radial velocity technique.  Their long periods also present a challenge to relatively short duration observations.}  Sensitivity extends down to Uranus- and Neptune-mass planets at separations of 20 to 30 au for the low viscosity condition ($\alpha$ = $10^{-4}$).  This is roughly consistent with the semi-major axes of Uranus and Neptune (19 and 30 au, respectively) in our own solar system, and suggests that such planets are detectable in existing observations of protoplanetary disks by model fitting directly in the visibility domain (assuming proper disk conditions).  It also overlaps many of the planet masses and separations that we detect in the medium viscosity case ($\alpha$ = $10^{-3}$), indicating that the sensitivity of this approach is broadly similar under conditions of low and medium disk viscosity.  Moreover, this mass-radius parameter space is not covered by existing exoplanet detection techniques and reinforces the analysis of disk substructure, under the planet-disk interaction hypothesis, as a complement to mature exoplanet surveys for gaining a better understanding of planetary system architectures (\citealt{Andrews_2020}).

For planets simulated at 5 au, model fitting was very sensitive to the number of Gaussians employed in the fitting routine and produced non-physical peaks and valleys in many of the derived radial intensity profiles.  This confused our detection test and we ultimately decided to exclude planet simulations at 5 au from consideration.  Defaulting to a single Gaussian fit invariably produced radial intensity plots without obvious substructure.  Higher resolution observations could help uncover such planets in future model-fitting campaigns.

In general terms, model fitting is most sensitive to massive planets at large relative separations.  Detection is aided by conditions of low or medium disk viscosity. Although not shown here, we also recover the gaps carved by a 11 $M_\oplus$ planet at 50 au for $\alpha = 10^{-3}$ and $10^{-4}$. Since the 11 $M_\earth$ planet is marginally detectable, we expect that Super-Earths below that mass may be difficult to be detected with the existing observations.

%----------editing by SZ-----------------------
\section{Young Planet Population \label{sec:5}}
The Taurus survey \citep{Long_2018, Long_2019} is an ideal sample to study the potential young planet population since it samples a full range of disk types around solar-mass stars with spectral type earlier than M3. The young planet population inferred is less biased than the DSHARP \citep{Andrews_2018b} and ODISEA \citep{Cieza_2021} surveys, where bright and large disks were preferentially selected. In this section we derive a population of young planet masses using the gap width in the Taurus sample.

\textit{Gap Width and Planet Mass.} We use the gap width normalized by the gap outer edge\footnote{We will just use ``gap width'' hereafter.} ($\Delta$) to infer potential young planet masses, as it is less prone to variation due to sensitivity than the gap depth (e.g., \citealt{Kanagawa_2016, Zhang_2018, Lodato_2019, Auddy_2020}). Generally speaking, the gap width is positively correlated with the planet mass. However, this relation can be complicated by the disk's gas and dust components. A stronger gas viscosity $\alpha$ makes the gap narrower. The dust-gas coupling is determined by the Stokes number, St, which is proportional to the dust size, a$_{\mathrm{max}}$ over the gas surface density. The coupling is stronger when St $\ll$ 1, and the gap is narrower; whereas when St $\sim$ 1, the dust drifts the fastest in the gas, so the gap can be much wider. \citet{Zhang_2018} measure this empirical relation between gap width and planet mass, under different viscosities and St. Then with the assumption of the disk parameters, we can infer the planet mass from the gap width.

\subsection{Planet Population}
\label{sec:5.2}

\begin{figure*}[t!]
\centering
\includegraphics[width=0.99\textwidth]{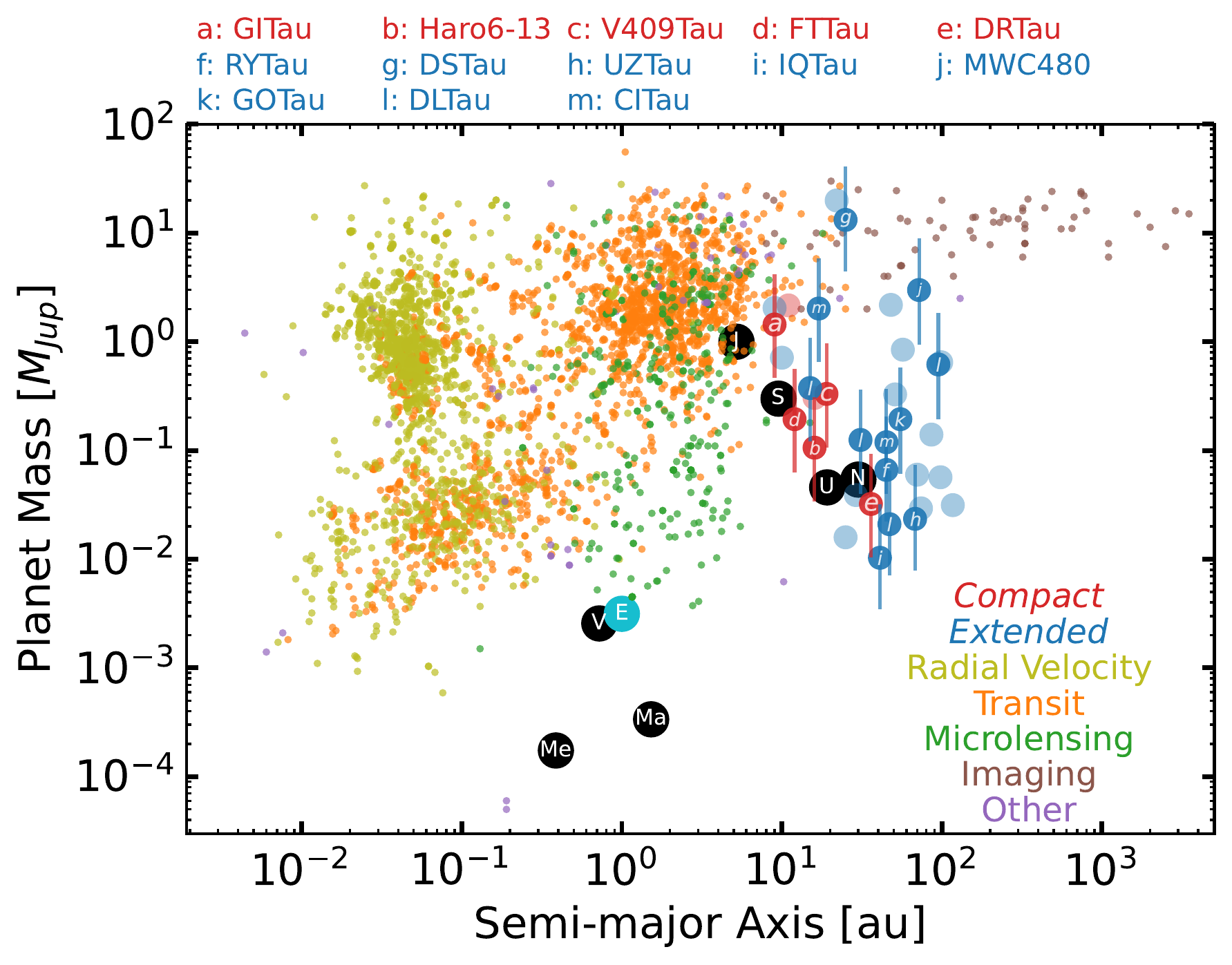}
\caption{Potential young planets in the Taurus sample with detected exoplanets on the planet mass-semi-major axis diagram. The planet mass is calculated as the mass assuming $a_{\mathrm{max}}$ $=$ 0.1 mm and $\alpha=10^{-3}$ (see Table \ref{tab:4} in the Appendix). The semi-major axis is equivalent to the gap location. Compact disks are in red and extended disks are in blue. The number inside the circle indicates the disk in which the planet resides. The error bar encompasses the uncertainties in the fitting and the disk viscosity. Exoplanets detected by various methods are marked with different colors.}
\label{fig:8}
\end{figure*}

In Figure \ref{fig:8}, we place potential planets derived from the Taurus survey on a planet mass-semi-major axis diagram, along with Solar System planets and confirmed exoplanets (from the \citet{NEA} as of April 12), in the same way as Figure 20 in \citet{Zhang_2018}. They are derived using the gap width, local dust surface density and an assumption of maximum grain size ($a_{\mathrm{max}}$) and disk viscosity ($\alpha$) \citep{Zhang_2018, Zhang_2022}. We plot the masses assuming a$_{\mathrm{max}}$ $=$ 0.1 mm and $\alpha=10^{-3}$. We have to assume these disk parameters since we do not have strong constraints on these parameters. Different assumptions of maximum grain size and disk viscosity would result in systematically different planet masses (an increase of disk viscosity, the dominant uncertainty, by a factor of ten would result in a planet calculated to be twice as massive). We also assume the planet is at the gap location. To be consistent with our injection-recovery study in Section \ref{sec:4} (Figure \ref{fig:3}), we exclude any planet below the detection limit. We also remove the planet for DL Tau at 66 au, since it can be the secondary gap carved by a single planet at 95 au, similar to the case in AS 209 \citep{Guzman_2018, Zhang_2018}. The complete mass estimation can be found in Table \ref{tab:4} in the Appendix, where we include all possible planets including those are excluded in the figure. We note that GO Tau D12 and FT Tau D26 are on the boundary of detection and non-detection zones. We choose to exclude them for the planet population and occurrence studies. Each disk is labeled with an alphabet representing its size in ascending order (a - e are compact disks and f - m are extended disks, where GI Tau being the smallest). The error bar accounts for the uncertainties in the fitting and a range of viscosities ($\alpha$ from 10$^{-4}$ to 10$^{-2}$). We also include the potential young planet population uncovered by the DSHARP survey using the same method with more transparent circles, also separated into compact and extended disks, but without alphabetical labels and error bars. We only include gaps with width ratios ($\Delta$) $>$ 0.12 in the figure since the uncertainties are too large to infer planet masses from the narrow gaps. However, we still list their tentative properties in the table. Plateaus (e.g., D5 in DO Tau) do not have a local minimum, so the method in \citet{Zhang_2018} cannot be applied, and they are not listed in the table, in line with Figures \ref{fig:6} and \ref{fig:7}. We also do not estimate the planet masses in the inner cavities of the transition disks in the Taurus sample, as their masses can vary throughout a large range or may even host multiple planets (e.g., there are at least two planets in PDS 70 \citep{Keppler_2018, Muller_2018, Wagner_2018, Haffert_2019, Isella_2019, Christiaens_2019,Wang_2020,Hashimoto_2020}). Compared to \citet{Zhang_2018}, we add the following gaps to the DSHARP sample: D25 of Elias 20, D29 of RU Lup, D98 of DoAr 25, and D117 of IM Lup, as their gap widths are also $>$ 0.12. Their masses are inferred using the fitting method, instead of by direct comparison with simulations. Exoplanets detected by various methods are represented by scatter points of different colors. The distribution of planets from the Taurus survey is similar to the distribution from the DSHARP survey.

\begin{figure}[t!]
\centering
\includegraphics[width=0.495\textwidth]{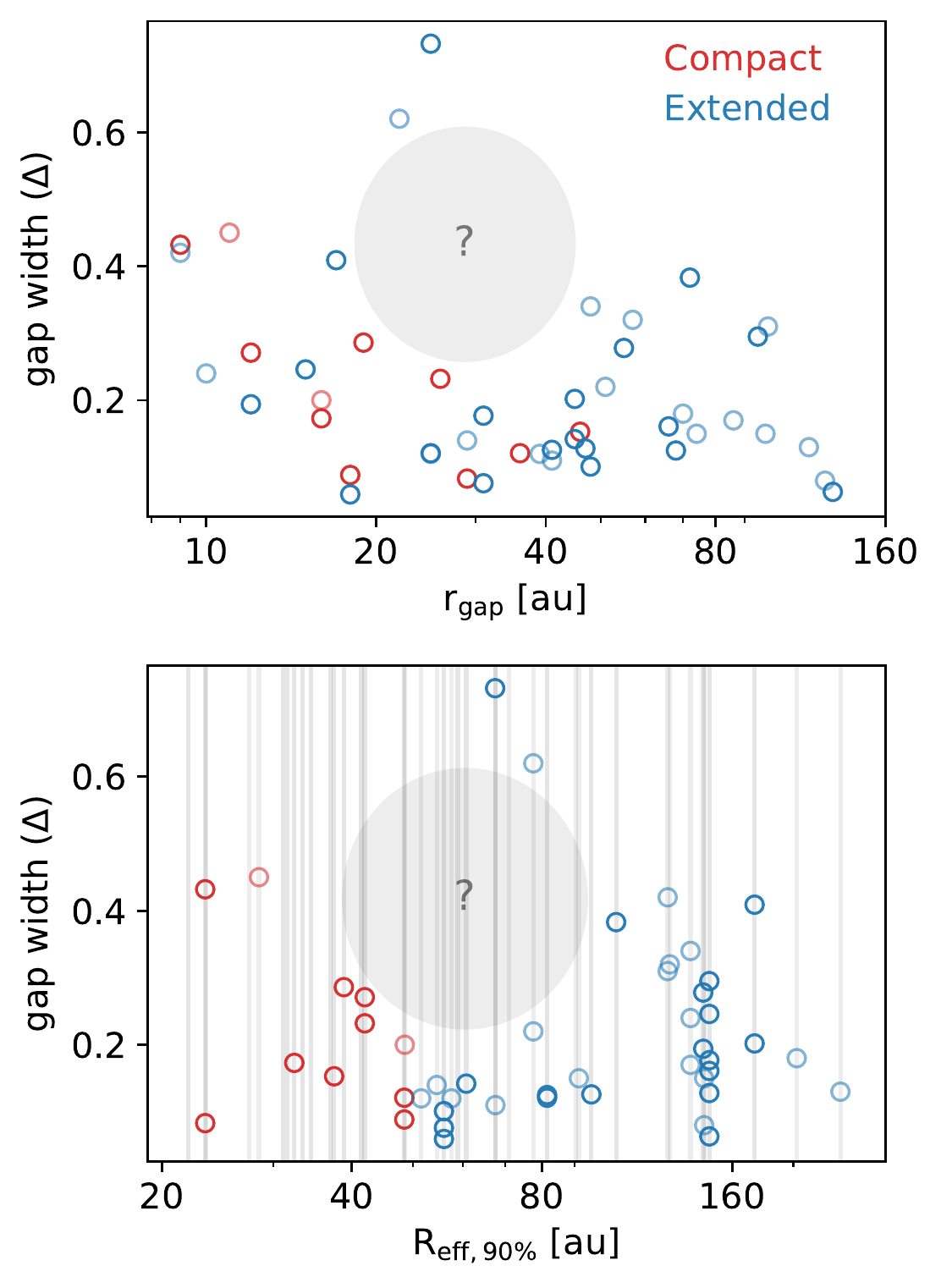}
\caption{The gap width vs. gap location (top panel) and gap width vs. effective disk radius (bottom panel) for Taurus and DSHARP samples. Gaps in compact disks are in red and extended disks in blue. The gaps in the DSHARP sample are marked in more transparent colors. For the bottom panel, each vertical line represents a disk radius. Points threaded by the same vertical line belong to the same disk. The intermediate regions with fewer wide gaps are marked by ellipses and question marks.}
\label{fig:8p}
\end{figure}

Interestingly, the giant planets ($>$ 0.1 $M_J$) between 20-50 au are fewer compared to smaller and larger separations. As planet masses are inferred from gap widths, this echoes our finding in Figure \ref{fig:6}, where wide gaps are rarer between 20-50 au. Since the definition of the gap width ($\Delta$) used to calculate planet mass is slightly different from the one in Figure \ref{fig:6} (normalized by the outer gap edge instead of the gap location), we remake the gap width-gap location plot in Figure \ref{fig:8p}'s top panel. It is similar to Figure \ref{fig:8} and also to Figure \ref{fig:6} except the x-axis is shown in log-scale. We mark the region with fewer wide gaps in ellipse and question mark. Additionally, this dearth of wide gaps is even more evident if we use the effective radius $\mathrm{R_{eff,90\%}}$ as the x-axis (bottom panel of Figure \ref{fig:8p}). For $\mathrm{R_{eff,90\%}}$ between 30-90 au, most of the gap widths are below 0.2.

More disk surveys of high angular resolution observations are needed to test out whether this void of wide gaps for intermediate sized disks (or gaps with intermediate separations) are statistically significant. If they are, it indicates that gaps at the inner and outer disks may be affected by different physical processes (e.g. different planet formation or evolution mechanisms).

\begin{figure}[t!]
\centering
\includegraphics[width=0.495\textwidth]{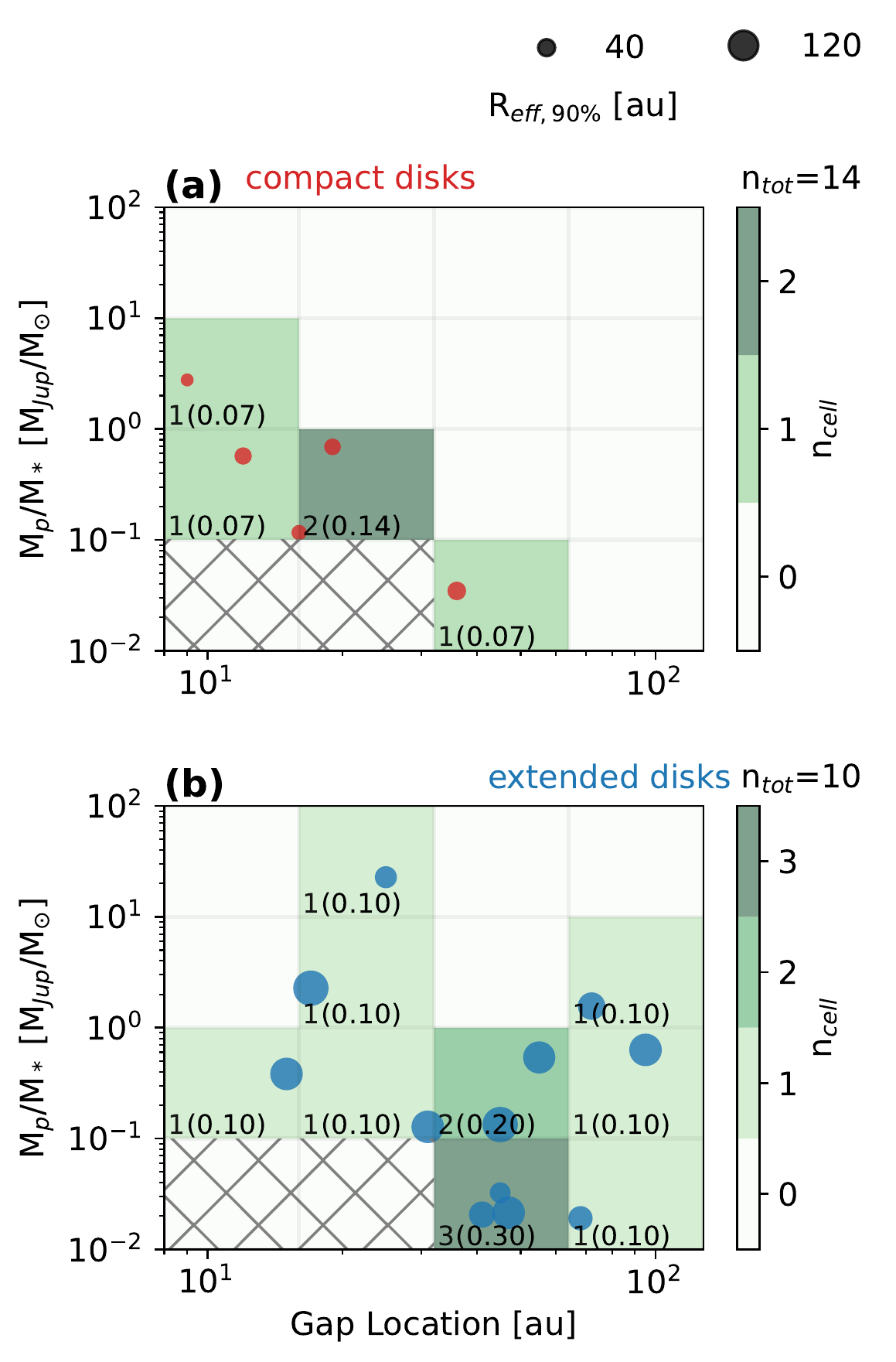}
\caption{Counts of potential young planets in the Taurus sample in each cell of a planet mass-gap location diagram, separated by compact (top panel) and extended (bottom panel) disks. The counts are indicated in the lower-left corner of each cell. The counts over the total number of compact or extended disks are listed in the parentheses. The hatched region is the detection limit. The y-axis is the planet-star mass ratio in units of Jupiter mass over solar mass. The disk size is proportional to the marker size, as indicated by the legend in the top right corner.}
\label{fig:9}
\end{figure}

In Figure \ref{fig:9}, we expand on the analysis by counting potential planets in mass-location bins among compact and extended disks in the Taurus survey.\footnote{For simplicity, we use M$_p$ to represent M$_p$/M$_*$ when we discuss Figures \ref{fig:9} and \ref{fig:10}.} The y-axes are the planet-star mass ratios in units of Jupiter mass over solar mass. We choose to present the planet-star mass ratio instead of the absolute planet mass since it directly reflects the observable, gap width. Another advantage of using the mass ratio is that we can directly compare our results with those from microlensing surveys. The gap locations are binned with edges at 8, 16, 32, 64, and 128 au, uniformly on a logarithmic scale. The planet masses are binned from 10$^{-2}$ M$_J$ to 100 M$_J$, with each bin spanning one decade in mass. The choice of bin sizes is arbitrary but able to cover all potential planets. The lower-left hatched region is the detection limit of our observations and model-fitting approach as summarized in Figure \ref{fig:3}.

In panels (a) and (b), we count the number of planets in the cell (indicated in the lower-left corner of each cell) and divide by the total number of the compact or extended disks in the Taurus subsample of 24 disks in single star systems. If the count is zero, the cell is empty. Both categories have most of their planets at the disk’s outer edges (20 to 30 au for the compact disks and 30 to 80 au for the extended disks). Since the planet location is assumed to be at the gap center, the planet location distribution is essentially the gap incidence distribution in Figure \ref{fig:5} with shallow gaps ($\Delta <$0.12) or low-mass planets below the detection limit removed. 
The most populated cell among compact disks is ([0.1, 1] $M_J$, [16, 32] au); the most populated cell among extended disks is ([0.01, 0.1] $M_J$, [32, 64] au). 

\section{Young Planet Occurrence \label{sec:6}}
With the young planet population, we calculate the planet occurrence rates on the planet-mass-semi-major-axis diagram. These occurrence rates are tentative due to the small sample sizes and several possible biases that are difficult to quantify. Still, they can be improved by future surveys with larger sample size and more homogeneous sample selections. We also include the DSHARP sample for comparison, but their biases are even larger. We introduce two kinds of occurrence rates with different assumptions in Section \ref{sec:6.1}, and list possible biases that affect the results in Section \ref{sec:6.2}. For the large differences in biases between Taurus and DSHARP samples, we calculate their occurrence rates separately. Then we present results and compare them with current exoplanet statistics in Sections \ref{sec:6.3} and \ref{sec:6.4}.
 
An earlier work by \citet{van_der_Marel_2021} calculated the occurrence rates of structured disks under different stellar masses using all known Class II disk surveys. With the assumption that all extended disks are structured and that all disks with substructures are due to giant planets, they found that current exoplanet demographics can account for all of the disk substructures. While \citet{van_der_Marel_2021} had a much larger disk sample size, the purpose of our exercise is to directly calculate the planet occurrence rates on the planet-mass-semi-major-axis diagram as exoplanet statistics, with planet masses ranging from sub-Neptunes to giant planets, and semi-major axis ranging from 8 au to 128 au.

\subsection{Two types of occurrence rates \label{sec:6.1}}
\begin{figure*}[t!]
\centering
\includegraphics[width=0.99\textwidth]{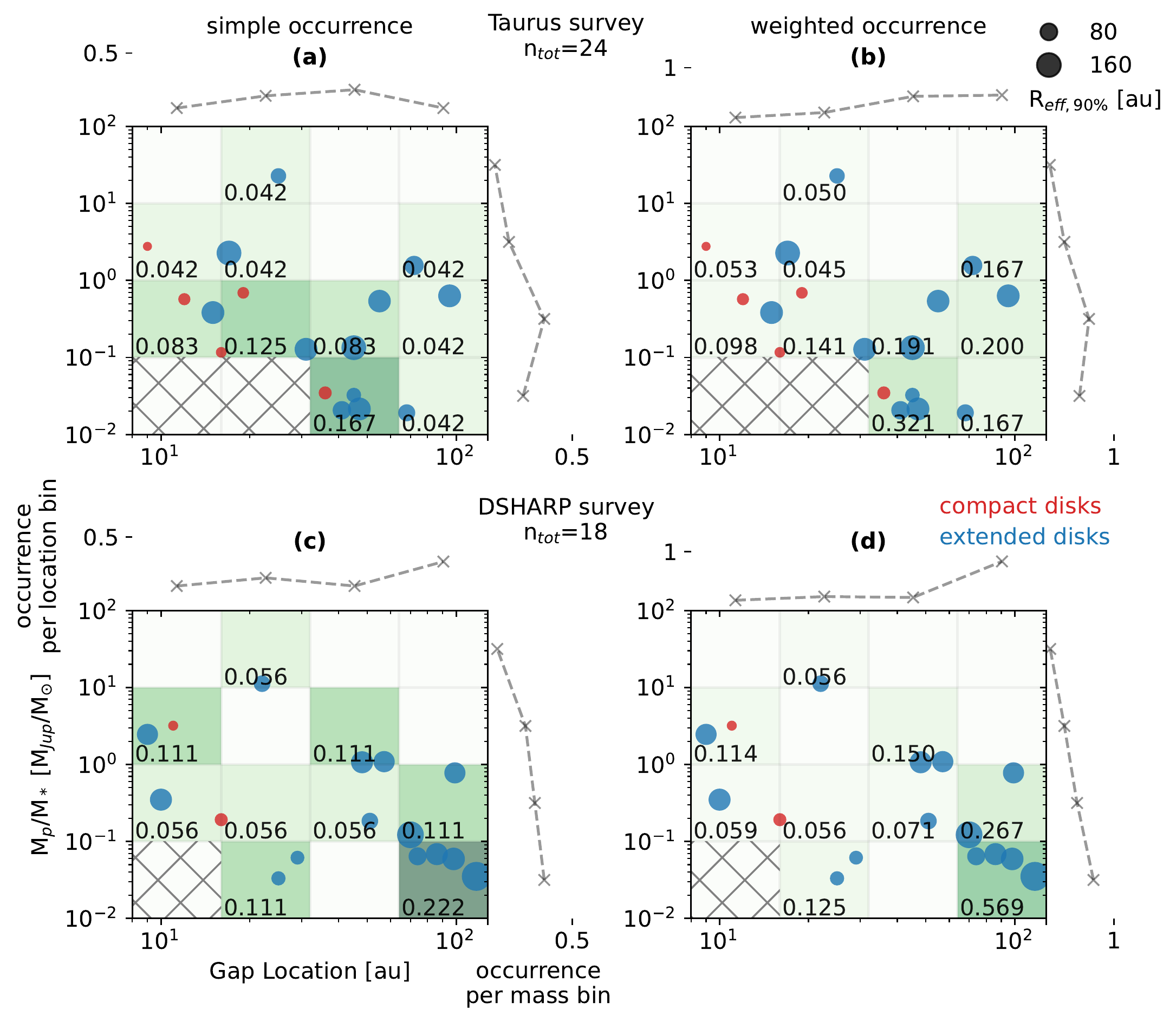}
\caption{The simple (left) and weighted (right) occurrence rates of the potential young planets in the Taurus (top) and DSHARP (bottom) samples. The fraction of planets relative to the total number of disks in the sample (or relative to the fraction of disks larger than the planet location and inner cavity's radius smaller than the planet location in the weighted case) is indicated in the lower left corner of each cell. The marginalized occurrence rates are indicated on the top and right of each histogram. Red circles are compact disks and blue circles are extended disks.}
\label{fig:10}
\end{figure*}

In Figure \ref{fig:10}, we use two different ways to correct for the bias and derive the planet occurrence rates for the Taurus and DSHARP samples (a version showing absolute planet masses is in the Appendix Figure \ref{fig:12}.).

We first use a simple way to calculate the occurrence rates (panels a and c), that is $n_{pl,cell}$/$n_{tot}$, where $n_{pl,cell}$ is the number of planets in a cell, and $n_{tot}$ is the total number of disks in a sample. This is based on the assumptions that (a) all planets form in the dusty disk; (b) whenever they are massive enough, they open gaps; (c) each gap detected corresponds to a planet in a disk; and (d) all planets have been detected. We denote this occurrence as ``simple occurrence''.

The second kind of occurrence rate is the weighted planet occurrence rate (panels b and d).
%, analogous to that used in exoplanet statistics \citep{Howard_2012}:

\begin{equation}
\label{eq:5}
\begin{aligned}
f_{cell} = \sum_{j=1}^{n_{pl,cell}} \frac{1}{n_{*,j}},
\end{aligned}
\end{equation}

\noindent where $n_{*,j}$ is the number of disks with radii larger than the radius of the gap and with radii of the inner cavity smaller than the radius of the gap, and $j$ goes over every potential planet in a cell. We use R$\textsubscript{eff,90\%}$ as the disk radius (if we use R$\textsubscript{eff,68\%}$ some substructures in the outer disk will be larger than the disk radius itself). We use the innermost ring's location as the cavity size. One assumption of the weighted rate (a) is that there are planets in the outer disk outside of the dusty disk (e.g., kinematic planets, \citealt{DiskDynamics_20}, or planets below the sensitivity limit, \citealt{Ilee_2022}). Since we cannot detect them, we need to account for these missing planets. Similarly, we also assume planets can be within the cavity where our gap width fitting method cannot be used. The second assumption is that (b) the planet's spatial distribution is independent of the disk size distribution and the cavity size distribution. Each detected planet will be corrected by (i.e., divided by) the fraction of disks larger than the planet location, and the fraction of disks with cavity sizes smaller than the planet location. We denote this occurrence as ``weighted occurrence'', which is typically larger than the simple occurrence.

We focus on the planet occurrence only at the Class II disk stage. In both types of occurrence rates, we assume that all stars should experience Class II disk stage. We also assume that the sample selection among the Taurus star-forming region is uniform among Class II disks.

\subsection{Possible Sources of Biases \label{sec:6.2}}

The young planet population is deduced with the assumption that all substructures are due to the planet. In this sense, these occurrence rates can be treated as upper limits, as other mechanisms can also possibly explain these substructures. Between two types of occurrences, simple occurrence is less than the weighted occurrence since it neglects planets in the disk's inner cavity and regions beyond the detected dusty disk.

\subsubsection{Detection Bias}

For the Taurus sample, we correct the observation completeness using the test results from Figure \ref{fig:3}, assuming $\alpha$ = 10$^{-3}$. 
As Figure \ref{fig:3} shows, disk viscosity, planet mass, and planet location can affect the detectability. Other than these factors, disk inclination, observation configuration, stellar luminosity, disk scale height, disk gas and dust surface density, and dust size also affect the detectability. The probability of detection is likely to be around 100\% for wide separation, high mass planets, and lower closer to the detection zone boundary. This means the biases of our results are higher for close separation, low mass planets. Ideally, similar to the radial velocity (RV) \citep{Mayor_2011} and transit surveys \citep{Howard_2012}, we could perform an injection-recovery study by varying all possible disk parameters and planet masses for a particular disk and use the specific ALMA observation configuration to simulate the synthetic observations. Then we do the parametric fitting for the suite of models to retrieve radial profiles. Finally, we calculate the possibility of the detection and factor that into the occurrence rate. We choose not to include this analysis in the current work due to our small sample size already teeming with large uncertainties. This process can be considered when we have a larger sample in the future.

For the DSHARP sample, we do not perform the parametric fitting since the DSHARP disks have much higher angular resolutions. We use the radial profiles from \citet{Andrews_2018b, Huang_2018} produced through standard \texttt{tclean} task. These radial profiles have similar resolutions to those derived from our parametric fitting among Taurus disks. As indicated by \citet{Jennings_2022a}, using super-resolution fitting barely increase the prominent substructures among this sample.

\subsubsection{Selection Bias}

The Taurus sample's disks are around stars of spectral type earlier than M3 \citep{Long_2019}. It excluded known binaries with separations between 0\arcsec.1 and 0\arcsec.5. It also excluded sources with high extinction or faint optical/near-IR emissions to avoid edge-on disks and embedded objects. These selections led to the missing of some low-mass disks. However, many of these low-mass disks are around close binaries. For them, the small size of the disk is due to tidal truncation. It is still possible that some of the low-mass disks are Class II disks around single stars. Unfortunately, it is difficult to quantify what fraction of the low-mass disks should have been counted as Class II disks around single stars. Since we do not make any correction for this fraction, we implicitly assume all the disks excluded are not Class II disks around single stars. The other possible large bias is that this sample selection excluded disks with existing high-resolution data, and these disks are most likely large and bright. This bias is also difficult to quantify. Although the Taurus sample is much more uniform than the DSHARP sample, it still biases towards disks with intermediate sizes and fluxes.

Quantifying the biases in the DSHARP sample is even more difficult. As the survey aimed to obtain as many substructures as possible \citep{Andrews_2018b}, bright full disks were selected and transition disks were avoided. The sample is also a combination of disks from several star-forming regions, so one would need to consider biases in each star-forming region's database on SED and mm dust continuum emissions that come into the sample selection.

The stellar masses in our samples range from 0.3 M$_\odot$ to 2 M$_\odot$ for Taurus and DSHARP samples. While the stellar mass is quite uniformly spaced within the mass range at least for the Taurus sample, the large span of the masses and the small number of disks can lead to large uncertainties if the occurrence rate is strongly dependent on the stellar mass. We note that the RV studies \citep{Mayor_2011, Fernandes_2019, Fulton_2021} and the \textit{Kepler} survey \citep{Thompson_2018} have much more controlled sample selection so the biases can be formally corrected. For example, the \textit{Kepler} sample has stellar mass centered on 1.0 M$_\odot$ with a standard deviation of 0.3 M$_\odot$ \citep{Huber_2017}.

\subsubsection{Other Assumption}

The calculated occurrence rates are the occurrence rates of young planets in systems with protoplanetary disks (Class II systems). We can also translate these occurrence rates to the lower limit of the young planet occurrence rates in the whole star forming region (including Class III disks) by multiplying the rates with the disk fraction $F$ (e.g., \citealt{Mulders_2021}). This approach intrinsically assumes that stars without disks do not have planets, which is why it is a lower limit. $F$ is strongly dependent on the age of the star-forming region. For example, \citet{Ribas_2014} (Table 4 therein) provide these fractions from mid-infrared observations between 22-24 $\mu$m in several star-forming regions ($F$=58\% for Taurus; 43\% for Lupus; 51\% for Ophiuchus; 22\% for Upper Sco).

\subsection{Occurrence Rate \label{sec:6.3}}
With all the caveats in mind, we can discuss the results. Aside from the occurrence rate in each grid, Figure \ref{fig:10} also shows the marginalized occurrence rates along gap locations and planet masses on the sides of the histogram. The simple occurrence rate for the Taurus sample peaks at region between 30-50 au, whereas the occurrene rates increases all the way beyond 100 au for the DSHARP sample. This can be due to the larger disk sizes in DSHARP sample. The planet occurrence rate decreases with increasing planet masses. The Taurus sample has a peak between 0.1-1 M$_J$, whereas the DSHARP sample has the largest occurrence rate between 0.01-0.1 M$_J$. This difference can be due to the higher detection limit for the Taurus sample. It can also due to the fact that stellar masses in the Taurus sample are on average lower than those in the DSHARP sample (see Table 1 in \citealt{Long_2019} and Table 1 in \citealt{Andrews_2018b}). Lower mass stars can have thicker disks (due to lower gravity), and thus have higher minimum gap-opening planet masses at a given location \citep{Sinclair_2020}. When adopting the weighted occurrence rate, the difference between two samples are smaller. Both of them have higher planet occurrence rates at the larger radii, and at or below a fraction of M$_J$. Owning to the small samples, the differences between two samples are not significant. By performing the two-sample Kolmogorov-Smirnov (KS) test using the \texttt{ks\_2samp} task in the PYTHON SCIPY package \citep{scipy}, we confirm that the marginalized occurrence distributions are
indistinguishable between Taurus and DSHARP samples for both location and mass distributions ($p$ = 0.77 for simple occurrence and $p$ = 1.0 for weighted occurrence). That both samples' occurrence rates peak at a planet-star mass ratio near the Neptune-Sun mass ratio is consistent with results from microlensing surveys \citep{Suzuki_2016b}.

Tentatively, we can also compare the Taurus and DSHARP planet occurrence rates with direct imaging, microlensing and radial velocity (RV) surveys. We report the occurrence rates binned either in planet masses or planet-star mass ratios depending on what is used in the previous literature.

To estimate the uncertainty of the rates, we assume the number of planets within a given range follows a binomial distribution \citep{Howard_2012}. They are drawn from an effective total number of stars (i.e., $n_{*,eff,cell}$ = $n_{pl,cell}$/$f_{cell}$). For the simple occurrence rate, $n_{*,eff,cell}$ is just the total number of disks in a survey. For the weighted occurrence rate, it is the effective (average) number of disks, which is less than or equal to the total number of disks. The quoted $\pm 1\sigma$ are at the 15.9 and 84.1 percentile levels in the cumulative distribution function.

\subsubsection{Direct Imaging}

The occurrence rate for wide orbit ($>$ 10 au) giant planets ($\sim$5 to 13 $M_J$) from the direct imaging method is believed to be around 1\% \citep{Bowler_2018}, with some studies suggesting it could be as high as 5 to 10\% \citep{Meshkat_2017, Nielsen_2019}. Here both samples have one planet as such, which give occurrence rates of 4 to 6\% (the simple and weighted occurrence rates are 4.2$\pm$4.2\% and 5.0$\pm$5.0\% for the Taurus sample, and 5.6$\pm$5.6\% and 5.6$\pm$5.6\% for the DSHARP sample, respectively), consistent with the direct imaging surveys.

\subsubsection{Microlensing}
Most of our parameter space (orbits $>$ 20 au and planet masses between sub-Neptune to Jupiter mass) is unique to the disk and cannot yet be probed by other detection methods, but substructures at $\sim$10 au start to overlap with planet locations from microlensing surveys. We compare our occurrence rates per decade in planet-star mass ratio ($q$) and separation with values from the microlensing survey of \citet{Suzuki_2016b}. First, we count our planets from all locations. Since these disk substructure locations span a little more than a decade (from 8 to 128 au), we divide our marginalized values (in Figure \ref{fig:10}) by 1.2 to be consistent with their definition. The simple and weighted occurrence rates for the Taurus and DSHARP samples are summarized in Table \ref{tab:2}. \citet{Suzuki_2016b} find the peak of the occurrence rate at $q$ = 1.7$\times 10^{-4}$ (or 0.18 $M_J$/$M_\odot$, 3.3 $M_{Nep}$/$M_\odot$) to be 61$^{+21}_{-16}$\%. It is similar to our highest weighted occurrence values for both samples, but we do not have enough resolution to compare at a specific narrow mass bin. Their median occurrence rate between 0.01 and 0.1 $M_J$/$M_\odot$ is around 20\%, which is consistent with our simple occurrence rates in both samples. Their median occurrence rate between 0.1 and 1 $M_J$/$M_\odot$ is also around 20\%. Our simple occurrence rates are also consistent with the result, whereas the weighted occurrence rates are much higher. To compare with the parameter space that overlaps microlensing surveys, we also calculate the occurrence rates between 8 and 16 au, normalized as per logarithmic radius. For planet mass ratios from 0.1 to 1 $M_J/M_\odot$, occurrence rates are 41.5$\pm$27.7\% for simple occurrence (48.2$\pm$31.6\% for weighted occurrence) for the Taurus sample and 36.9$\pm$18.5\% for simple occurrence (38.0$\pm$19.5\% for weighted occurrence) for the DSHARP sample. These are higher than the microlensing results, but consistent within 1$\sigma$. For planet mass ratios from 1 to 10 $M_J/M_\odot$, rates are 13.8$\pm$13.8\% for simple occurrence (17.5$\pm$17.5\% for weighted occurrence) for the Taurus sample and 36.9$\pm$18.5\% for simple occurrence (38.0$\pm$19.5\% for weighted occurrence) for the DSHARP sample, higher than those in the microlensing survey.

\begin{table}[t!]
\centering
\textbf{Table 2} \\
\text{Planet Occurrence Rates for Taurus and DSHARP Samples} \\
\smallskip
 \begin{tabular*}{0.4725\textwidth}{c @{\extracolsep{\fill}} cccc}
 \hline \hline
 Mass Ratio & Survey & Simple & Weighted \\
 ($M_J/M_\odot$) &  & Occurrence & Occurrence \\
 \hline
    $[$0.01, 0.1) & Taurus & 17.4$\pm$8.3\% & 40.7$\pm$20.0\% \\
    $[$0.01, 0.1) & DSHARP & 27.8$\pm$11.1\% & 57.9$_{-11.1}^{+22.2}$\%  \\
    $[$0.1, 1) & Taurus & 27.8$\pm$8.3\% & 52.4$\pm$15.4\% \\
    $[$0.1, 1) & DSHARP & 23.1$\pm$11.1\% & 37.7$\pm$18.2\%  \\
    $[$1, 10)  & Taurus &  10.4$\pm$8.3\% & 22.1$_{-18.2}^{+9.1}$\% \\
    $[$1, 10)  & DSHARP & 18.5$\pm$11.1\% & 22.0$\pm$13.3\%  \\
    $[$10, 100) & Taurus & 3.5$\pm$3.5\% & 4.2$\pm$4.2\% \\
    $[$10, 100) & DSHARP & 4.6$\pm$4.6\% & 4.6$\pm$4.6\%  \\
 \hline
 \end{tabular*}
\rtask{tab:2}
\end{table}

\citet{Poleski_2021} calculate the occurrence rate for ice giants (planets with orbits from 5 to 15 au and mass ratios from 0.01 to 3.3 $M_J/M_\odot$) to be 1.4$^{+0.9}_{-0.6}$ per microlensing star, 2.4$\sigma$ higher than what was found in \citet{Suzuki_2016b}. Our occurrence rates estimated from substructures beyond 8 au are lower. The simple (and weighted) occurrences are 12.5$\pm$8.3\% (15.0$\pm$10.0\%) for the Taurus survey and 16.7$\pm$11.1\% (17.3$_{-11.8}^{+5.9}\%$) for the DSHARP survey, much lower than what is found in \citet{Poleski_2021}. On the other hand, Neptune-mass planets within 15 au might still be hidden under the detection limits.

\begin{figure*}[t!]
\centering
\includegraphics[width=0.99\textwidth]{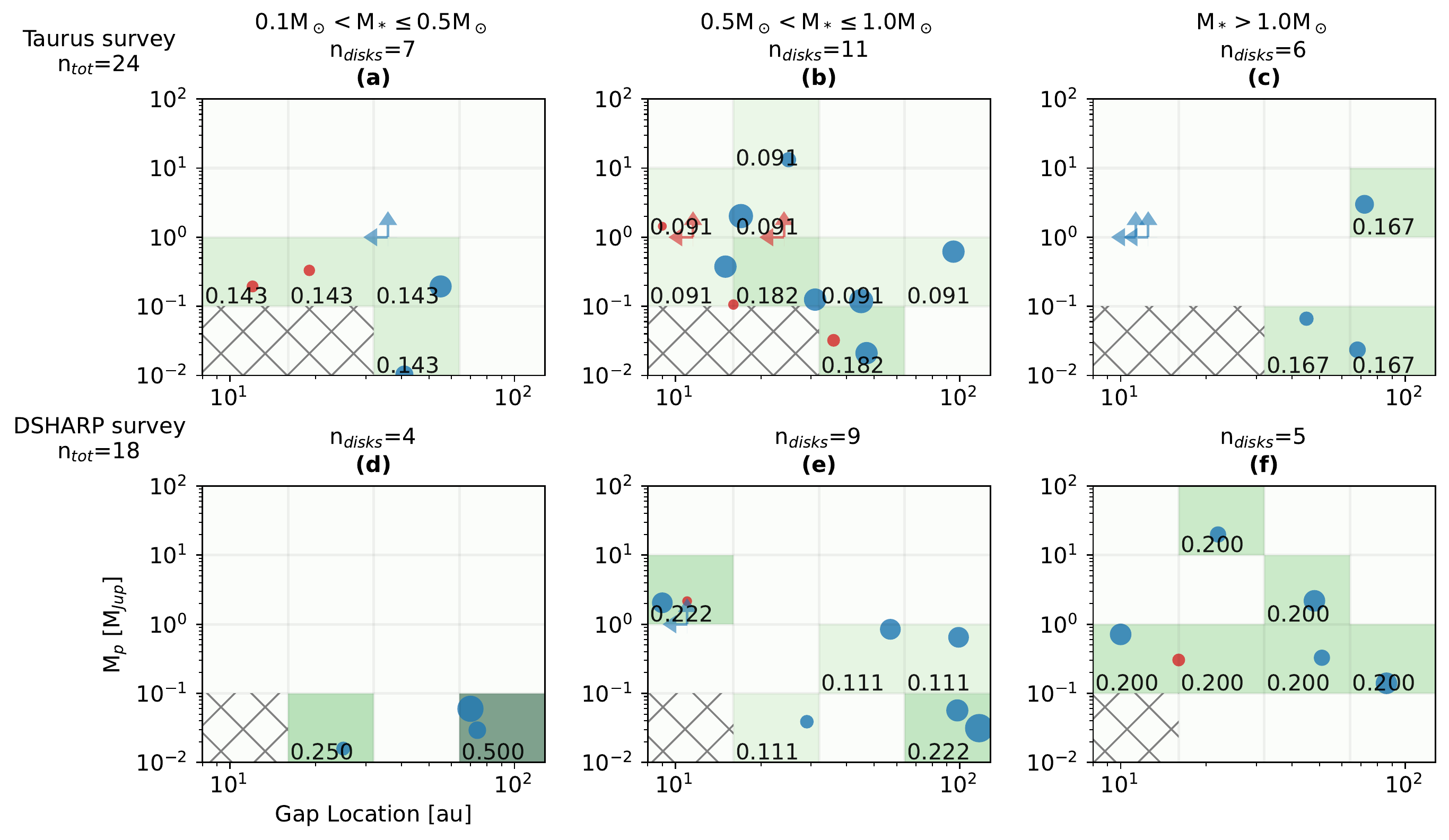}
\caption{The simple occurrence rates of the potential young planets in the Taurus and DSHARP samples separated by three different stellar mass bins, with boundaries at 0.5 and 1 M$_{\odot}$. The layout for each diagram is similar to Figure \ref{fig:10}, but the y-axis represents the absolute planet mass. The number of disks that have host stellar masses within a certain range is marked at the top of each diagram. The arrows are five potential giant planets in the transition disks in the Taurus survey. We use the ring location outside the cavity as the upper limit of the planet location and one Jupiter mass as the lower limit of the planet mass, since the planet locations and masses are both uncertain. Given these uncertainties, and the fact that multiple planets can exist in a cavity, these potential giant planets are not included in the occurrence rates.}
\label{fig:11}
\end{figure*}

\subsubsection{Radial Velocity (RV)}
%A contradictory result is from an extrapolation of RV surveys. 
\citet{Fernandes_2019} indicate that the occurrence rate of giant planets (masses between 0.1 to 20 $M_J$ and 0.1 to 100 au from their host stars) is 26.6\%, using a fitting with a broken power law and a turnover at the snowline on existing data within several astronomical units. \citet{Fulton_2021} corroborate the broken power-law with 2$\sigma$ significance but found a shallower decline beyond the snowline. \citet{Fernandes_2019} infer that the disk substructure occurrence rate is much higher than that, so only part of the substructure can be due to planets. We calculate the simple and weighted occurrences to be 50$\pm$8.3\% and 94.4$\pm$7.7\% for the Taurus survey and 55.6$\pm$11.1\% and 79.7$_{-7.7}^{+15.4}$\% for the DSHARP survey. These are higher than the extrapolated value in \citet{Fernandes_2019}. However, the simple occurrence for the Taurus sample is still consistent with their value within 3$\sigma$. As such, the hypothesis that all gaps are due to planets cannot be ruled out solely based on their extrapolated occurrence rate. Other than non-planet origins for disk gaps, the slight discrepancy can be explained if the decrease in occurrence rate beyond snowlines cannot be extrapolated, or that these samples are somehow biased toward higher substructure occurrence rates. Perhaps it is also possible that the young planet population is very different from the mature population, as young planets will continue to grow and migrate \citep{Nayakshin_2022}. To compare directly with current exoplanet statistics, these samples could be used as an initial condition for population synthesis studies to evolve them to a mature stage \citep{Wang_2022}.

\subsection{Occurrence for different stellar types \label{sec:6.4}} 

The dependency between planet occurrence and stellar mass can also be tested using planets inferred from disk gaps. In Figure \ref{fig:11}, we separate the Taurus (top panels) and DSHARP (bottom panels) samples into three stellar mass regimes: M dwarfs (0.1 M$_{\odot}$ $<$ M$_*$ $\leq$ 0.5 M$_{\odot}$), Sun-like stars (0.5 M$_{\odot}$ $<$ M$_*$ $\leq$ 1 M$_{\odot}$) and massive stars (M $_*$ $>$ 1 M$_{\odot}$).  The y-axis is the absolute planet mass in Jupiter mass. We only calculate the simple occurrence rates, which are the number of planets in a cell divided by the number of disks within the given mass range (marked as $n_{disks}$). There are five transition disks in the Taurus survey that could also affect the occurrence rates. However, since their masses, locations and numbers are all uncertain, we only mark them using arrows in the figure and choose not to include them in the calculations of the occurrence rates. The upper limit of the planet location is taken as the outer ring of the cavity, whereas the lower limit of their planet masses is set at 1 M$_J$. With our calculated occurrence rates, we can compare the Taurus and DSHARP samples to trends found in exoplanet surveys.

While the stellar mass dependence from \textit{Kepler} is mainly for sub-Neptunes within 1 au, we can test out their trends for our samples on much larger separations. From the \textit{Kepler} survey, sub-Neptune exoplanets are more common around low-mass stars  \citep{Howard_2012, Mulders_2018}. Their occurrence rate around M stars is a factor of 2 to 4 higher than their occurrence rate around FGK stars \citep{Mulders_2015c, Gaidos_2016}. Our results show that the occurrence rates for sub-Neptune planets around M dwarfs, Sun-like stars and massive stars are 14.3\%$\pm$14.3\%, 27.3$\pm$9.1\% and 16.7$\pm$16.7\% for the Taurus sample, and 50.0$\pm$25.0\%, 22.2$\pm$11.1\% and 0\% for the DSHARP sample. For the DSHARP survey, the factor of $\sim$2 higher occurrence rates around M dwarfs relative to Sun-like stars is consistent with the \textit{Kepler} survey. However, we need higher resolution observations of compact disks to study sub-Neptune planets in the inner disks.

Giant planets occur more frequently around high-mass stars \citep{Johnson_2010}, but the dependence is weaker and less statistically significant than the dependence with metallicity \citep{Mulders_2018}. A caveat is that the dependence of giant planets with stellar mass includes many planet below a Jupiter mass \citep{Ghezzi_2018, Fulton_2021}. The occurrence rates of planets $>$ 1 M$_J$ among the three mass regimes are 0\%, 27.3$\pm$9.1\% and 16.7\%$\pm$16.7\% for the Taurus sample, and 0\%, 22.2$\pm$11.1\% and 40.0$\pm$20.0\% for the DSHARP sample. Direct imaging surveys constrain the occurrence rate of planets $>$ 1 M$_J$ and $>$ 10 au separation to be 5.7\%$^{+3.8}_{-2.8}$ for FGK stars \citep{Vigan_2017, Vigan_2021}. The values calculated here are much higher. \citet{Nielsen_2019} find the occurrence rate for $M_* >$ 1.5 M$_\odot$ to be 24$_{-10}^{+13}$\%, while \citet{Vigan_2021} find 23.0$_{-9.7}^{+13.5}$\% for BA stars. Our values are consistent with these results within 1$\sigma$. Perhaps the most relevant result comes from the NaCo-ISPY large program, where young planets were searched around 45 young stars surrounded by protoplanetary disks \citep{Cugno_2023}. Most of the stars in the survey have $M_* >$ $M_{\odot}$. The occurrence rates for semis-major axis between 20-500 au, and $T_{\mathrm{eff}}$ between 600-3000 K are 21.2$_{-13.6}^{+24.3}$\%, 14.8$_{-9.6}^{+17.5}$\%,
10.8$_{-7.0}^{+12.6}$\% for R$_p$ = 2, 3, 5 R$_J$. While the masses of these young planets are difficult to constrain, our occurrence rates are roughly consistent with these values if both populations have similar planet masses.

% \citet{van_der_Marel_2021} calculated the occurrence rates of structured disks under different stellar masses using all known Class II disk surveys. With the assumption that all extended disks are structured and that all disks with substructures are due to giant planets, they find that current exoplanet demographics can account for all of the disk substructures.

%----------editing by SZ-----------------------

\section{Summary}
\label{sec:7}

We reanalyze 1.33 mm ALMA continuum observations of 12 compact disks in the Taurus star-forming region, initially classified as smooth and featureless in the analysis of \citet{Long_2019}. We also reanalyze 12 disks with substructures found \citet{Long_2018}. Our approach is to fit their deprojected visibility profiles directly, which is more sensitive to small-scale disk substructure. We compare the incidence and characteristics of substructure in the populations of compact and extended disks within the Taurus sample of 24 disks in single star systems and the high-resolution DSHARP survey. We then test the sensitivity of these observations and our fitting approach by analyzing mock observations of substructures created by planet-disk interaction simulations. We conclude by calculating potential planet masses and occurrence rates and comparing them to the results of mature exoplanet surveys.  Our main findings are as follows:

\begin{enumerate}
  \item We detect substructure in the continuum emission of 6 out of 12 compact disks in our sample. This at least doubles the number of compact disks ($R\textsubscript{eff,90\%}$ $<$ 50 au) with known substructure. Detected substructures are generally shallow and narrow, with widths on order of 5 au.  On average, the 6 smooth disks are both fainter and more compact than the six disks with identified substructure.
  \item No substructures are detected in compact disks with $R\textsubscript{eff,90\%}$ below 23 au or continuum luminosities below 12 mJy.  Assuming such disks are not universally featureless, this highlights the need for better spatial resolution observations of the most compact disks.
  \item Combining our work with the results from \citet{Long_2018} and the DSHARP survey (\citealt{Huang_2018}), we find that substructure is present in two-thirds of compact disks ($R\textsubscript{eff,90\%}$ $<$ 50 au).  Our statistics indicate that the occurrence of substructure peaks near 20 au for compact disks and 40 au for extended disks, but this could be an artifact of biases against substructure detection near disk center and in faint outer regions.  Substructure is otherwise present out to large relative radii.  There may be no preferred location for substructure in either compact or extended disks.
  \item In the Taurus and DSHARP samples, gap widths and depths are greater in extended disks than in compact disks, but this discrepancy largely disappears after normalizing gap width to gap location.
  \item Tentatively, we find that fewer wide gaps are located between 20-50 au or found within disks with $R\textsubscript{eff,90\%}$ between 30-90 au. If all of the gaps are created by planets, this means that fewer giant planets exist at intermediate separations or in intermediate sized disks.
   \item Given the case of low or medium disk viscosity $(\alpha \le 10^{-3})$, our fitting approach detects Neptune-mass planets at separations of 20 to 30 au.  This mass-radius parameter space is not probed by existing exoplanet detection techniques, and suggests that substructure analysis could be used to improve our understanding of system architectures under the planet-disk interaction hypothesis.
  \item With our newly detected substructures, we infer more potential planets within compact disks and obtain a more complete view of the young planet population.
  \item We calculate planet occurrence rates for both the Taurus and DSHARP surveys using simple and weighted methods.  Occurrence rates in both samples roughly increase with separation from the host star, and decrease with planet mass.  Planet occurrence is highest for Neptune-mass planets, and in the outer disks.  The values from both samples are consisent within 1$\sigma$ in each bin.
  \item In a large parameter space, our calculated occurrence rates for the Taurus sample broadly agree with microlensing and direct imaging surveys.  For 0.01 $M_J/M_\odot$ $\lesssim$ $M_p/M_*$ $\lesssim$ 0.1 $M_J/M_\odot$, the rate is 17.4$\pm$8.3\%, consistent with microlensing surveys.  For 0.1 $M_J/M_\odot$ $\lesssim$ $M_p/M_*$ $\lesssim$ 1 $M_J/M_\odot$, it is 27.8$\pm$8.3\%, higher than the results from microlensing surveys but consistent within 1$\sigma$.  For gas giants more massive than 5 $M_J$, the occurrence rate is 4.2$\pm$4.2\%, which is consistent with direct imaging surveys.
\end{enumerate}

% SZ: basically echoing what's added in the abstract
% With newly detected substructures, we can infer more potential young planets within 40 au and obtain a more complete view of the young planet population.

% We calculate the occurrence rates for both the Taurus and DSHARP surveys using simple and weighted methods. The Taurus and DSHARP sample share great similarities. Both samples have occurrences increasing with separations, but decreasing with planet masses. The occurrence is the highest around Neptune mass planets and at the outer disks. The values in two samples are consistent within 1$\sigma$ in each bin.

% In a large parameter space, the occurrence rates agree with direct imaging, microlensing and RV surveys.
% For the Taurus sample it is 17.4$_{-8.3}^{+8.3}$\%, for 0.01 $M_J/M_\odot$ $\lesssim$ $M_p/M_*$ $\lesssim$ 0.1 $M_J/M_\odot$, consistent with microlensing surveys.
% The occurrence is 20.8$_{-8.3}^{+8.3}$\%, for 0.1 $M_J/M_\odot$ $\lesssim$ $M_p/M_*$ $\lesssim$ 1 $M_J/M_\odot$, higher than the microlensing surveys.
% The occurrence rate for gas giants $>$ 5$M_J$ is 4.2$^{+4.2}_{-4.2}$\%,, consistent with direct imaging surveys.

\section*{Acknowledgements}

This paper makes use of the following ALMA data: ADS/JAO.ALMA\#2016.1.01164.S.  ALMA is a partnership of ESO (representing its member states), NSF (USA) and NINS (Japan), together with NRC (Canada), MOST and ASIAA (Taiwan), and KASI (Republic of Korea), in cooperation with the Republic of Chile.  The Joint ALMA Observatory is operated by ESO, AUI/NRAO and NAOJ.  The National Radio Astronomy Observatory is a facility of the National Science Foundation operated under cooperative agreement by Associated Universities, Inc.  This work has made use of data from the European Space Agency (ESA) mission {\it Gaia} (\url{https://www.cosmos.esa.int/gaia}), processed by the {\it Gaia} Data Processing and Analysis Consortium (DPAC, \url{https://www.cosmos.esa.int/web/gaia/dpac/consortium}).  Funding for the DPAC has been provided by national institutions, in particular the institutions participating in the {\it Gaia} Multilateral Agreement.  This research has made use of the NASA Exoplanet Archive, which is operated by the California Institute of Technology, under contract with the National Aeronautics and Space Administration under the Exoplanet Exploration Program.

We thank the referee for valuable comments that significantly improved the quality of the paper. S.Z. and Z.Z. acknowledge support from NASA through the NASA FINESST grant 80NSSC20K1376.  S.Z. acknowledges support from Russell L. and Brenda Frank Scholarship. Z.Z. acknowledges support from the National Science Foundation under CAREER grant AST-1753168 and support from NASA award 80NSSC22K1413. Support for F.L. was provided by NASA through the NASA Hubble Fellowship grant \#HST-HF2-51512.001-A awarded by the Space Telescope Science Institute, which is operated by the Association of Universities for Research in Astronomy, Incorporated, under NASA contract NAS5-26555. K.Z., M.K, and L.T. acknowledges the support of the Office of the Vice Chancellor for Research and Graduate Education at the University of Wisconsin – Madison with funding from the Wisconsin Alumni Research Foundation.

\section*{Appendix}

\begin{figure*}[t!]
\centering
\includegraphics[width=0.7\textwidth]{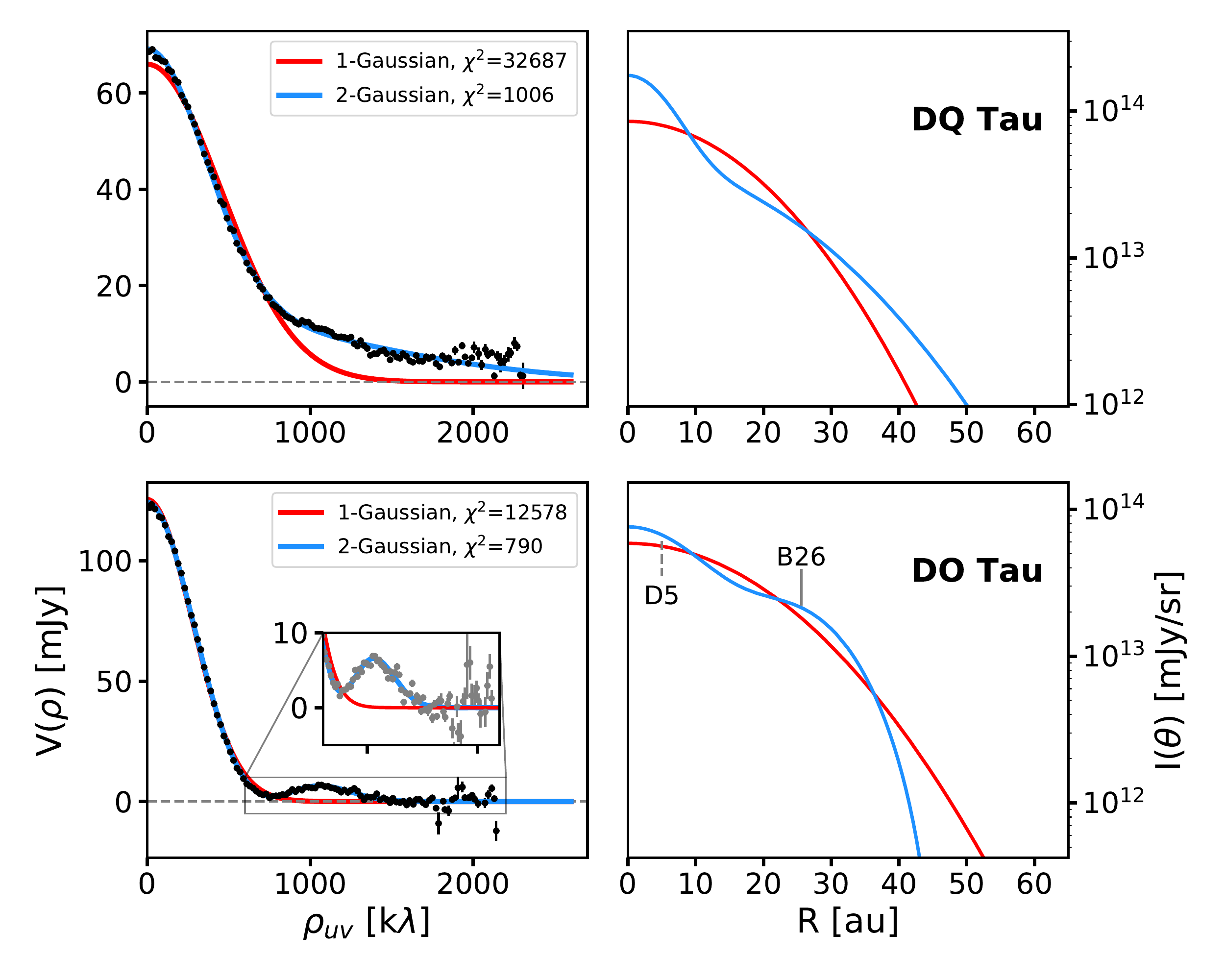}
\vspace{-0.3cm}
\caption{Comparison between 1 and 2 Gaussian-components models for DQ Tau and DO Tau disks. It shows that the 1-Gaussian models largely under-reproduce the visibilities beyond 1000\,k$\lambda$, while the 2-Gaussian models reproduce the data at longer baselines much better. \label{fig:dqtau_comparison}}
\label{fig:14}
\end{figure*}

\begin{figure*}[t!]
\centering
\includegraphics[width=0.99\textwidth]{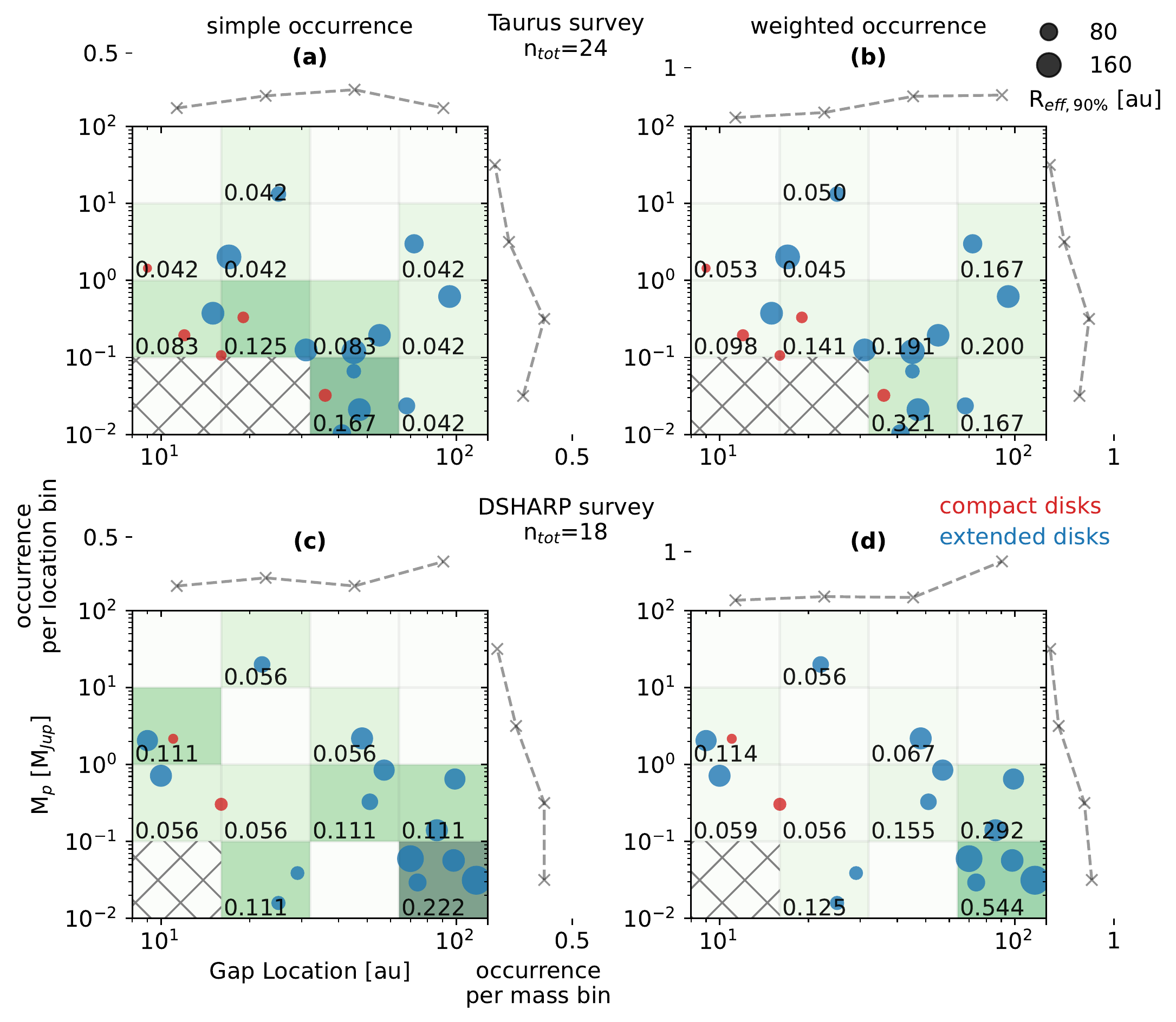}
\caption{Same as Figure \ref{fig:10}, but the y-axes are planet masses rather than planet-star mass ratios.}
\label{fig:12}
\end{figure*}

Table \ref{tab:3} shows the derived disk parameters from \citet{Long_2019} that were used in this work.  For example, distance is used to scale to a standard luminosity, disk flux is used to make parameter space comparisons, effective radius is fundamental in the determination of compact and extended disks, and inclination and position angle are used in visibility deprojections. Table \ref{tab:1p} shows the best-fit parameters in Equation \ref{eq:4} and chi-square values for these 12 compact disks. Figure \ref{fig:14} demonstrates that two-Gaussian component models of DO Tau and DQ Tau reduce the chi-square values by an order of magnitude compared to that of one-Gaussian models. Thus, we adopt two-Gaussian models for these two disks.  Table \ref{tab:4} shows the planet masses inferred from all Taurus gaps, including those narrow and/or tentative gaps excluded from the figures.  Figure \ref{fig:12} is a recreation of Figure \ref{fig:10}, but with absolute planet masses on the y-axis instead of the planet-star mass ratio. 

\begin{figure*}[t!]
\centering
\includegraphics[width=0.99\textwidth]{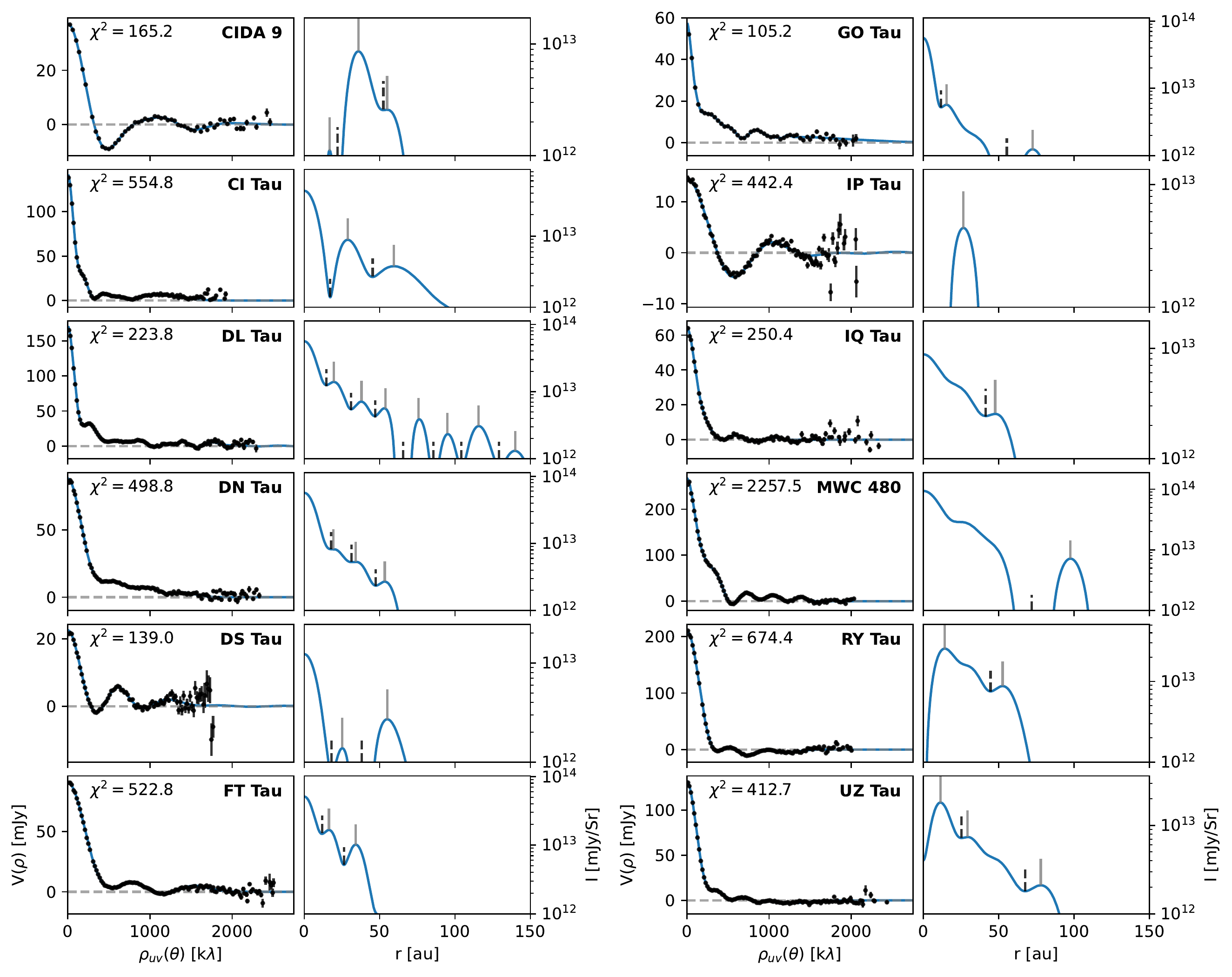}
\caption{Deprojected visibility and radial intensity profiles for the 12 disks with identified substructures in \citep{Long_2018} following our fitting approach.  Overlaid in blue on the visibility curves are our best-fit models, which are used to derive the adjacent radial intensity profiles. The $\chi^2$ score of the fitting is marked on each panel. Dashed black lines on the radial intensity curves of panel (a) mark gaps, and solid gray lines mark rings. These feature names can be found in Table \ref{tab:5a}.}
\label{fig:13}
\end{figure*}

Figure \ref{fig:13} shows the reanalysis of 12 disks with substructures found in \citet{Long_2018}, along with the substructures listed in Table \ref{tab:5a}. Among these 12 disks, DL Tau and FT Tau have the most gaps and rings within extended and compact disks, respectively. DL Tau has 7 gaps and 7 rings. We note that the spacing of the gaps resembles that of AS 209. Since one planet is possible to carve out multiple gaps with characteristic spacing when $\alpha$ is low \citep{Zhu_2014, Bae_2017, Dong_2017, Dong_2018, Bae_2018a, Bae_2018b, Zhang_2018, Miranda_2019}, a planet at 95 au may also explain the secondary gap at 66 au, or even at 47 au \citep{Zhang_2018}. Thus, we only use one planet to explain the ring at 95 au and the gap at 66 au in the planet population and occurrence rate calculation. FT Tau has 2 gaps and 2 rings, implying that compact disks might also be highly substructured if we observe them with a higher resolution.

\begin{table*}[t!]
\centering
\textbf{Table 3} \\
\text{Compact Disk Model Parameters} \\
\smallskip
 \begin{tabular*}{\textwidth}{c @{\extracolsep{\fill}} ccccc}
 \hline \hline
 Name & Distance (pc) & $F_\nu$ (mJy) & $R\textsubscript{eff,90\%}$ (au) & Inclination (deg) & PA (deg) \\
 \hline
 BP Tau & 129 & $45.15^{+0.19}_{-0.14}$ & 37.49 & $38.2^{+0.5}_{-0.5}$ & $151.1^{+1.0}_{-1.0}$ \\
 DO Tau & 139 & $123.76^{+0.17}_{-0.27}$ & 33.43 & $27.6^{+0.3}_{-0.3}$ & $170.0^{+0.9}_{-0.9}$ \\
 DQ Tau & 197 & $69.27^{+0.15}_{-0.19}$ & 37.02 & $16.1^{+1.2}_{-1.2}$ & $20.3^{+4.3}_{-4.3}$ \\
 DR Tau & 195 & $127.18^{+0.20}_{-0.22}$ & 48.42 & $5.4^{+2.1}_{-2.6}$ & $3.4^{+8.2}_{-8.0}$ \\
 GI Tau & 130 & $17.69^{+0.25}_{-0.07}$ & 23.43 & $43.8^{+1.1}_{-1.1}$ & $143.7^{+1.9}_{-1.6}$ \\
 GK Tau & 129 & $5.15^{+0.19}_{-0.11}$ & 11.61 & $40.2^{+5.9}_{-6.2}$ & $119.9^{+8.9}_{-9.1}$ \\
 Haro 6-13 & 130 & $137.10^{+0.24}_{-0.21}$ & 32.42 & $41.1^{+0.3}_{-0.3}$ & $154.2^{+0.3}_{-0.3}$ \\
 HO Tau & 161 & $17.72^{+0.20}_{-0.17}$ & 41.39 & $55.0^{+0.8}_{-0.8}$ & $116.3^{+1.0}_{-1.0}$ \\
 HP Tau & 177 & $49.33^{+0.16}_{-0.15}$ & 22.00 & $18.3^{+1.2}_{-1.4}$ & $56.5^{+4.6}_{-4.3}$ \\
 HQ Tau & 158 & $3.98^{+0.08}_{-0.17}$ & 31.09 & $53.8^{+3.2}_{-3.2}$ & $179.1^{+3.2}_{-3.4}$ \\
 V409 Tau & 131 & $20.22^{+0.12}_{-0.18}$ & 38.84 & $69.3^{+0.3}_{-0.3}$ & $44.8^{+0.5}_{-0.5}$ \\
 V836 Tau & 169 & $26.24^{+0.16}_{-0.12}$ & 31.58 & $43.1^{+0.8}_{-0.8}$ & $117.6^{+1.3}_{-1.3}$ \\
 \hline
 \vspace{-2.5mm}
 \end{tabular*}
{\raggedright \textbf{Notes:} This table is a recreation of Tables 1 and 3 from \citet{Long_2019}, with new effective radii calculated from the model intensity profiles produced in this work.  Distance estimates are from Gaia DR2 parallax data (\citealt{Gaia_2016, Gaia_2018}).  Distances and effective radii are shown without uncertainties as those uncertainties are very small (1\% or less). \par}
\rtask{tab:3}
\end{table*}

\begin{table*}[t!]
\centering
\textbf{Table 4} \\
\text{Compact Disk Best-fitting Gaussian Parameters} \\
\smallskip
 \begin{tabular*}{\textwidth}{c @{\extracolsep{\fill}} cccc}
 \hline \hline
Disk   & a$_i$                  & $\sigma_i$         & $\rho_i$    & $\chi^2$ \\
 \hline
 BP Tau &                        &                    &             & 317.2    \\
       & 1.126$\times$10$^{-1}$$\pm$3.54$\times$10$^{-2}$  & 1.377$\times$10$^{-1}$$\pm$8.5$\times$10$^{-2}$ & 0$\pm$0     &          \\
       & -4.72$\times$10$^{-2}$$\pm$1.05$\times$10$^{-1}$ & 1.654$\times$10$^{-1}$$\pm$7.4$\times$10$^{-2}$ & 425$\pm$307 &          \\
       & -2.19$\times$10$^{-2}$$\pm$1.65$\times$10$^{-3}$ & 2.294$\times$10$^{-1}$$\pm$3.9$\times$10$^{-2}$ & 1166$\pm$8  &          \\
 \hline
 DO Tau    &                                    &                                 &                          & 789.93   \\
          & 4.382$\times$10$^{-1}$$\pm$4.21$\times$10$^{-4}$  & 1.153$\times$10$^{-1}$$\pm$8.9$\times$10$^{-5}$ & 0$\pm$0     &          \\
          & 1.032$\times$10$^{-1}$$\pm$1.16$\times$10$^{-3}$  & 1.515$\times$10$^{-1}$$\pm$1.5$\times$10$^{-3}$ & 958$\pm$2   &          \\
 \hline
DR Tau    &                                    &                                 &                          & 1248.56  \\
          & 4.873$\times$10$^{-1}$$\pm$8.02$\times$10$^{-4}$  & 1.090$\times$10$^{-1}$$\pm$6.3$\times$10$^{-4}$ & 0$\pm$0     &          \\
          & 2.043$\times$10$^{-1}$$\pm$1.45$\times$10$^{-3}$  & 1.245$\times$10$^{-1}$$\pm$2.5$\times$10$^{-3}$ & 905$\pm$4   &          \\
          & 1.169$\times$10$^{-1}$$\pm$2.71$\times$10$^{-3}$  & 1.653$\times$10$^{-1}$$\pm$2.4$\times$10$^{-3}$ & 1646$\pm$5  &          \\
 \hline
 GI Tau    &                                    &                                 &                          & 483.54   \\
          & 8.36$\times$10$^{-2}$$\pm$1.41$\times$10$^{-3}$  & 9.03$\times$10$^{-2}$$\pm$8.9$\times$10$^{-4}$ & 0$\pm$0     &          \\
          & 5.31$\times$10$^{-2}$$\pm$4.32$\times$10$^{-3}$  & 9.05$\times$10$^{-2}$$\pm$6.4$\times$10$^{-3}$ & 1467$\pm$21 &          \\
 \hline
 Haro 6-13 &                                    &                                 &                          & 605.46   \\
          & 6.132$\times$10$^{-1}$$\pm$2.18$\times$10$^{-2}$  & 5.07$\times$10$^{-2}$$\pm$3.7$\times$10$^{-4}$ & 0$\pm$0     &          \\
          & -5.167$\times$10$^{-1}$$\pm$2.82$\times$10$^{-2}$ & 1.201$\times$10$^{-1}$$\pm$2.3$\times$10$^{-3}$ & 504$\pm$7   &          \\
          & 4.16$\times$10$^{-2}$$\pm$6.35$\times$10$^{-3}$  & 2.269$\times$10$^{-1}$$\pm$8.3$\times$10$^{-3}$ & 0$\pm$0     &          \\
 \hline
 V409 Tau  &                                    &                                 &                          & 391.08   \\
          & 5.91$\times$10$^{-2}$$\pm$3.63$\times$10$^{-4}$  & 1.446$\times$10$^{-1}$$\pm$6.2$\times$10$^{-4}$ & 0$\pm$0     &          \\
          & 2.66$\times$10$^{-2}$$\pm$1.02$\times$10$^{-3}$  & 1.636$\times$10$^{-1}$$\pm$5.4$\times$10$^{-3}$ & 804$\pm$6   &          \\
\hline
 DQ Tau  &                                    &                                 &                          & 1006.16   \\
          & 2.378$\times$10$^{-1}$$\pm$1.23$\times$10$^{-3}$ & 9.23$\times$10$^{-2}$$\pm$2.4$\times$ 10$^{-4}$ & 0$\pm$0     &          \\
          & 2.085$\times$10$^{-1}$$\pm$9.47$\times$10$^{-4}$ & 2.69$\times$10$^{-2}$$\pm$3.4$\times$10$^{-4}$ & 0$\pm$0   &          \\
\hline
 GK Tau  &                                    &                                 &                          & 417.22   \\
          & 4.93$\times$10$^{-2}$$\pm$ 4.75$\times$10$^{-4}$ & 4.20$\times$10$^{-2}$$\pm$5.2$\times$10$^{-4}$ & 0$\pm$0     &          \\
 \hline
HO Tau  &                                    &                                 &                          & 579.53   \\
          & 5.97$\times$10$^{-2}$$\pm$1.86$\times$10$^{-4}$ & 1.199$\times$ 10$^{-1}$$\pm$4.7$\times$10$^{-4}$ & 0$\pm$0     &          \\
\hline
 HP Tau  &                                    &                                 &                          & 1072.76   \\
          & 3.430$\times$ 10$^{-1}$$\pm$3.46$\times$10$^{-4}$ & 5.80$\times$10$^{-2}$$\pm$7.3$\times$10$^{-5}$ & 0$\pm$0     &          \\
\hline
 HQ Tau  &                                    &                                 &                          & 570.69   \\
          & 1.79$\times$10$^{-2}$$\pm$2.19$\times$10$^{-4}$ & 9.17$\times$10$^{-2}$$\pm$1.4$\times$10$^{-3}$ & 0$\pm$0     &          \\
\hline
 V836 Tau  &                                    &                                 &                          & 923.88   \\
          & 1.223$\times$10$^{-1}$$\pm$2.45$\times$ 10$^{-4}$ & 8.71$\times$10$^{-2}$$\pm$2.2$\times$10$^{-4}$ & 0$\pm$0     &          \\

 \hline
 \vspace{-2.5mm}
 \end{tabular*}
{\raggedright \textbf{Notes:}  $a_i$, $\sigma_i$, and $\rho_i$ are amplitudes, widths and central locations of Gaussians. The first Gaussian is always centered on zero, so its uncertainty is also fixed to zero. The last column shows the  $\chi^2$ values of the best-fitting. \par}
\rtask{tab:1p}
\end{table*}

% appendix --editing by SZ-------
% Table for inferred planet masses from the Taurus survey.
\begin{table*}[t!]
\centering
\textbf{Table 5} \\
\text{Planet Masses Inferred from Gaps in the Taurus Sample} \\
\smallskip
 \begin{tabular*}{\textwidth}{c @{\extracolsep{\fill}} cccccccccc}
 \hline \hline
 Disk & $M_*$ & Gap & Width & $M_{p,1em3,0.1mm}$ & $M_{p,1em3,1mm}$ & $M_{p,1em3,1cm}$ & error$_{0.1mm}$ &  error$_{1mm}$ & error$_{1cm}$ \\
     & ($M_{\odot}$) & (au) & $\frac{r_{out}-r_{in}}{r_{out}}$ & ($M_J$) & ($M_J$) & ($M_J$) & (dex) &  (dex) & (dex) \\
 \hline
   CI Tau & 0.89   &    17 &     0.41   &          2.02   &        1.63   &        0.767  & $^{+0.13}_{-0.16}$ & $^{+0.14}_{-0.17}$ &  $^{+0.21}_{-0.5}$ \\
          &        &    45 &     0.20   &          0.119  &        0.077  &        0.0086 & $^{+0.16}_{-0.14}$ & $^{+0.22}_{-0.16}$ & $^{+0.53}_{-0.63}$ \\
   DL Tau & 0.98   &    15 &     0.25   &          0.375  &        0.304  &        0.201  & $^{+0.13}_{-0.16}$ & $^{+0.14}_{-0.17}$ &  $^{+0.21}_{-0.5}$ \\
          &        &    31 &     0.18   &          0.125  &        0.101  &        0.0834 & $^{+0.13}_{-0.16}$ & $^{+0.14}_{-0.17}$ &  $^{+0.21}_{-0.5}$ \\
          &        &    47 &     0.13   &          0.0210 &        0.0156 &             - & $^{+0.16}_{-0.14}$ & $^{+0.22}_{-0.16}$ &                  - \\
          &        &    66 &     0.16   &          0.0531 &        0.0368 &             - & $^{+0.16}_{-0.14}$ & $^{+0.22}_{-0.16}$ &                  - \\
          &        &    95 &     0.30   &          0.618  &        0.362  &             - & $^{+0.14}_{-0.17}$ &  $^{+0.21}_{-0.5}$ &                  - \\
          &        &   129 &     0.06   &               - &             - &             - &                  - &                  - &                  - \\
   DN Tau & 0.52   &    18 &     0.06   &          0.0016 &             - &             - & $^{+0.13}_{-0.16}$ &                  - &                  - \\
          &        &    31 &     0.08   &          0.0039 &        0.0013 &             - & $^{+0.13}_{-0.16}$ & $^{+0.22}_{-0.16}$ &                  - \\
          &        &    48 &     0.10   &               - &             - &             - &                  - &                  -  &                  - \\
          
   DS Tau & 0.58   &    25 &     0.73   &         13.2    &        2.60   &             - & $^{+0.16}_{-0.14}$ &  $^{+0.21}_{-0.5}$ &                  - \\
   GO Tau & 0.36   &    12 &     0.19   &          0.0646 &        0.0267 &        0.0029 & $^{+0.13}_{-0.16}$ & $^{+0.22}_{-0.16}$ & $^{+0.53}_{-0.63}$ \\
          &        &    55 &     0.28   &          0.194  &        0.118  &             - & $^{+0.14}_{-0.17}$ &  $^{+0.21}_{-0.5}$ &                  - \\
   IQ Tau & 0.50   &    41 &     0.13   &          0.0103  &        0.0077 &             - & $^{+0.16}_{-0.14}$ & $^{+0.22}_{-0.16}$ &                  - \\
  MWC 480 & 1.91   &    72 &     0.38   &          2.99   &        0.271  &             - & $^{+0.14}_{-0.17}$ & $^{+0.53}_{-0.63}$ &                  - \\
   RY Tau & 2.04   &    45 &     0.14   &          0.0662  &        0.0965 &             - & $^{+0.16}_{-0.14}$ & $^{+0.21}_{-0.5}$ &                  - \\
   UZ Tau & 1.23   &    25 &     0.12   &          0.0409 &        0.0331 &        0.0352 & $^{+0.13}_{-0.16}$ & $^{+0.14}_{-0.17}$ &  $^{+0.21}_{-0.5}$ \\
          &        &    68 &     0.13   &          0.0235 &        0.0401 &             - & $^{+0.16}_{-0.14}$ &  $^{+0.21}_{-0.5}$ &                  - \\
       \hline
        FT Tau & 0.34 & 12 & 0.27 & 0.194 & 0.157 & 0.097 & $^{+0.13}_{-0.16}$ & $^{+0.14}_{-0.17}$ &  $^{+0.21}_{-0.5}$ \\
               &      & 26 & 0.23 & 0.0807 & 0.0502 & 0.0059 & $^{+0.16}_{-0.14}$ & $^{+0.22}_{-0.16}$ &  $^{+0.53}_{-0.63}$ \\
       \hline
BP Tau & 0.52 & 46 & 0.15 & 0.0016 & - & - & $^{+0.53}_{-0.63}$ & - & - \\
DR Tau & 0.93 & 18 & 0.089 & 0.0107 & 0.0087 & 0.0114 & $^{+0.13}_{-0.16}$ & $^{+0.14}_{-0.17}$ & $^{+0.21}_{-0.50}$ \\
       &        & 36 & 0.12 & 0.0322 & 0.0118 & - & $^{+0.13}_{-0.16}$ & $^{+0.22}_{-0.16}$ & - \\
GI Tau & 0.52 & 9 & 0.43& 1.44 & 0.726 & 0.115 & $^{+0.13}_{-0.16}$ & $^{+0.22}_{-0.16}$ & $^{+0.53}_{-0.63}$ \\
        &       & 29 & 0.08 & - & - & - & - & - & - \\
Haro 6-13 & 0.91 & 16 & 0.17 & 0.106 & 0.0858 & 0.0717 & $^{+0.13}_{-0.16}$ & $^{+0.14}_{-0.17}$ & $^{+0.21}_{-0.50}$ \\
V409 Tau & 0.48 & 19 & 0.29 & 0.331 & 0.151 & - & $^{+0.13}_{-0.16}$ & $^{+0.22}_{-0.16}$ & - \\
 \hline
 \vspace{-2.5mm}
 \end{tabular*}
{\raggedright \textbf{Notes:} All of the inferred planet masses from gaps in the Taurus sample as long as a non-zero value can be calculated following \citet{Zhang_2018}. Gaps with width ratios $<$ 0.12 are not shown in our plotted figures as uncertainties are too large to infer planet masses from the narrow gaps. The gap in BP Tau is tentative so the planet mass is also not shown in our plotted figures. Horizontal lines delineate extended disks and compact disks in \citet{Long_2018}, and compact disks in the current paper (from top to bottom). \par}
\rtask{tab:4}
\end{table*}

\begin{table}[t!]
\centering
\textbf{Table 6} \\
\text{Characteristics of Identified Substructures} \\
\text{(Taurus Disks in \citet{Long_2018})} \\
\smallskip
 \begin{tabular*}{0.4725\textwidth}{c @{\extracolsep{\fill}} cccc}
 \hline \hline
 Disk & Feature & $r_0$ (au) & Width (au) & Depth ($\frac{I_{d}}{I_{b}}$) \\
 \hline
 CI Tau& D17     & 17.2  & 9.2        & 0.157         \\
       & B29     & 28.9    & 14.4       & -             \\
       & D45     & 45.4  & 10.3       & 0.711         \\
       & B60     & 59.5  & 16.9         & -             \\
        \hline
CIDA 9A  & B17     & 16.8  & 1.4        & -             \\
       & D22     & 22.2  & 30.3       & 0.116         \\
       & B36     & 35.9  & 12.1       & -             \\
       & D53     & 52.5  & 2.41       & 1.000         \\
       & B55     & 55    & 2.44       & -             \\
        \hline
DL Tau  & D15     & 14.7  & 4.2        & 0.857         \\
       & B20     & 19.7  & 4.8        & -             \\
       & D31     & 31.1  & 6.1        & 0.726         \\
       & B38     & 38    & 7          & -             \\
       & D47     & 47.2  & 6.4        & 0.732         \\
       & B54     & 53.9  & 5.9        & -             \\
       & D66     & 65.5  & 12.6       & 0.250         \\
       & B76     & 75.8  & 7.9        & -             \\
       & D86     & 85.8  & 10.1       & 0.417         \\
       & B95     & 95    & 7.5        & -             \\
       & D104    & 104.2 & 12.6       & 0.323         \\
       & B116    & 115.5 & 11         & -             \\
       & D129    & 129.3 & 10.9       & 0.769         \\
       & B140    & 140   & 8.1        & -             \\
        \hline
DN Tau & D18     & 17.9  & 1.2        & 1.000            \\        
       & B19     & 19.3  & 1.2        & -             \\
       & D31     & 31.3  & 2.5        & 1.000         \\
       & B34     & 34.1  & 2.5        & -             \\
       & D48     & 47.5  & 5.1        & 0.852         \\
       & B53     & 53.4  & 5.5        & -             \\
        \hline
DS Tau  & D18     & 18.2  & 7.2        & 0.714         \\
       & B25     & 25.2  & 5.2        & -             \\
       & D38     & 38.2  & 35.6       & 0.370         \\
       & B55     & 55.1  & 12.9       & -             \\
        \hline
 \vspace{-2.5mm}
 \end{tabular*}
%{\raggedright \textbf{Notes:} \par}
\rtask{tab:5a}
\label{tab:5a}
\end{table}

\begin{table}[t!]
\centering
\textbf{Table 6} \text{(continued)}
\text{Characteristics of Identified Substructures} \\
\text{(Taurus Disks in \citet{Long_2018})} \\
\smallskip
 \begin{tabular*}{0.4725\textwidth}{c @{\extracolsep{\fill}} cccc}
 \hline \hline
 Disk & Feature & $r_0$ (au) & Width (au) & Depth ($\frac{I_{d}}{I_{b}}$) \\
 \hline
FT Tau  & D12     & 11.8  & 3.9        & 0.882         \\
       & B16     & 16.4  & 4.4          & -             \\
       & D26     & 26.4  & 6.9        & 0.520         \\
       & B34     & 34.1  & 8.0        & -             \\
        \hline
GO Tau  & D12     & 11.8  & 2.7        & 0.912         \\
       & B15     & 15.3  & 3.8        & -             \\
       & D55     & 55.4  & 26.2       & 0.833         \\
       & B73     & 72.5  & 6.6        & -             \\
        \hline
IP Tau  & B27     & 26.6  & 12.5       & -             \\
        \hline
IQ Tau  & D41     & 41.4  & 5.6        & 0.923         \\
       & B48     & 47.7  & 5.7        & -             \\
        \hline
MWC 480 & D72     & 72    & 35         & 0.139         \\
       & B98     & 97.6  & 13.7       & -             \\
        \hline
RY Tau  & B14     & 14.2  & 17.6       & -             \\
       & D45     & 44.6  & 6.9        & 0.852         \\
       & B53     & 52.6  & 7.5        & -             \\
        \hline
UZ Tau E  & B11     & 11.3  & 10.2       & -             \\
       & D25     & 25.3  & 3.4        & 0.973         \\
       & B29     & 29.4  & 3.9        & -             \\
       & D68     & 67.5  & 9          & 0.857         \\
       & B78     & 77.9  & 9.5        & -             \\
       \hline
 \vspace{-2.5mm}
 \end{tabular*}
%{\raggedright \textbf{Notes:} \par}
\rtask{tab:5b}
\end{table}

\clearpage

% appendix --editing by SZ-------

\end{document}